\documentclass[journal=jacsat,manuscript=article]{achemso}
\usepackage{graphicx} % Required for inserting images
\usepackage{subfigure} % Required for inserting two images
\usepackage{chemformula} % Formula subscripts using \ch{}
\usepackage[T1]{fontenc} % Use modern font encodings
\usepackage{xr} % For cross referencing
\SectionNumbersOn % For numbering the sections
\usepackage[normalem]{ulem}
\usepackage{multirow,booktabs,tabularray}
\usepackage{adjustbox}
\usepackage{colortbl}
\usepackage{array}
\usepackage{pdfpages}
\newcolumntype{V}[1]{>{\color{violet}}#1}

\title[An \textsf{achemso} demo]
{Long-Lived Isomers of \ch{C11H9+}: New Experimental Insights from the PIRENEA Setup} 
  
\author{Ana I. Lozano}
\author{Anthony Bonnamy}
\affiliation[Universit\'e de Toulouse]
{Institut de Recherche en Astrophysique et Plan\'etologie (IRAP), Universit\'e de Toulouse, CNRS, CNES, 9 Avenue du Colonel Roche, F-31028 Toulouse, France}
\author{Aude Simon}
\affiliation[Universit\'e de Toulouse]
{Laboratoire de Chimie et Physique Quantiques LCPQ/FeRMI, Universit\'e de Toulouse, CNRS, 118 Route de Narbonne, 31062 Toulouse, France}
\author{Christine Joblin}
\email{christine.joblin@cnrs.fr}
\affiliation[Universit\'e de Toulouse]
{Institut de Recherche en Astrophysique et Plan\'etologie (IRAP), Universit\'e de Toulouse, CNRS, CNES, 9 Avenue du Colonel Roche, F-31028 Toulouse, France}

\abbreviations{FTICR,MPD,MeNp,TDDFT,CC2}

\begin{document}

%%%%%%%%%%%%%%%%%%%%%%%%%%%%%%%%%%%%%%%%%%%%%%%%%%%%%%%%%%%%%%%%%%%%%
%% The abstract 
%%%%%%%%%%%%%%%%%%%%%%%%%%%%%%%%%%%%%%%%%%%%%%%%%%%%%%%%%%%%%%%%%%%%%
\begin{abstract}

The dehydrogenated cation of methylated benzene is known to exist in two isomeric forms: benzylium and tropylium. Structurally similar forms have been proposed for the -H cations of methylated polycyclic aromatic hydrocarbons, but their spectroscopic characterization remains limited, and their photophysical properties are still poorly understood.
Previous studies identified 2-naphthylmethylium and benzyltropylium as specific long-lived isomers of the -H fragment of methylnaphthalene cations.
Here, we investigate the photodissociation spectroscopy and photoprocessing of gas-phase \ch{C11H9+} ions in the visible range.
Experiments are conducted using the versatile laboratory astrophysics setup PIRENEA, which enables studies over long timescales ($\sim$1000~s) and allows photoprocessing to be combined with ion–molecule reaction experiments.
We confirm the presence of 2-naphthylmethylium and benzyltropylium and additionally identify 1-naphthylmethylium, previously undetected in experiments.
Moreover, we present the first complete quantitative analysis of the relative abundances of these isomers and provide clear evidence of interconversion among the three long-lived species. These isomerization processes occur below the dissociation threshold, a finding supported by molecular dynamics simulations. The photophysical properties of \ch{C11H9+} isomers --including isomerization and fluorescence-- make them intriguing candidates for consideration in astrophysical environments exposed to mild UV irradiation (h$\nu$ < 7~eV). Moreover, their detection via rotational spectroscopy in such regions is facilitated by their closed-shell electronic structure.

\end{abstract}


%%%%%%%%%%%%%%%%%%%%%%%%%%%%%%%%%%%%%%%%%%%%%%%%%%%%%%%%%%%%%%%%%%%%%
%% Introduction
%%%%%%%%%%%%%%%%%%%%%%%%%%%%%%%%%%%%%%%%%%%%%%%%%%%%%%%%%%%%%%%%%%%%%

\section{Introduction}
\label{Introduction}

Polycyclic aromatic hydrocarbons (PAHs) and their derivatives are major pollutants in urban air and aquatic environments \cite{Rochman2013, StMary2021, Sadiktsis2023}. PAHs formed through combustion and pyrolysis processes primarily consist of small (2–4 ring) species, along with less abundant larger compounds. Methylated PAHs, such as methylnaphthalene (\ch{C10H7-CH3}), are also significant, with their relative abundance to parent PAHs (e.g., naphthalene) varying based on precursor composition and combustion or pyrolysis conditions—an effect notably observed in the pyrolysis of waste plastics \cite{Zhou2015,Zhou2016}. Additionally, 1-methylnaphthalene is recognized as a key representative bicyclic aromatic compound in transportation fuels, highlighting the need to investigate its chemical kinetics under conditions relevant to internal combustion engines \cite{Wang2010,Liang2024}.

PAHs are also prominent constituents of astrophysical environments and are considered the leading candidates for the carriers of the aromatic infrared bands (AIBs), intense mid-infrared emission features. As AIB emission is triggered by ultraviolet (UV) photons, these bands are particularly prominent in photodissociation regions (PDRs), such as those found in massive star-forming regions. Studying them in relation to the photochemical evolution of their carriers is one of the key areas expected to benefit from James Webb Space Telescope observations in the coming years \cite{Peeters2024,Chown2024}.

Additionally, recent radioastronomical observations have revealed the presence of small aromatic molecules in the TMC-1 dark cloud and several other dark clouds \cite{Burkhardt2021a, Agundez2023}. Following the detection of cyano-benzene (\ch{C6H5CN}) \cite{McGuire2018}, subsequent discoveries have included indene (c-\ch{C9H8}), cyano-indene \cite{Cernicharo2021, Burkhardt2021b, Sita2022}, the two isomers (1- and 2-) of cyano-naphthalene (\ch{C10H7CN}) \cite{McGuire2021}, and, more recently, two isomers of cyano-acenaphthylene (\ch{C12H7CN}) \cite{Cernicharo2024} along with all three isomers of cyano-pyrene (\ch{C16H9CN}) \cite{Wenzel2024a, Wenzel2024b}. Due to the intrinsic limitations of radio detection, these observations primarily reveal molecules with significant dipole moments. However, they still confirm the presence of PAH derivatives containing two to four rings in dark molecular clouds. Similarly sized PAHs are also abundant in primitive Solar System materials, such as carbonaceous chondrites \cite{Sephton2002} and the C-type asteroid Ryugu \cite{Aponte2023, Zeichner2024, Sabbah2024}, further highlighting their widespread occurrence and significance. In these samples, the PAH population is primarily composed of small PAHs, some of which exhibit a significant degree of alkylation \cite{Basile1984,Plows2003, Elsila2005, Lecasble2022}.

The presence of methylated PAHs in photodissociation regions (PDRs) has been proposed to explain the 3.4~$\mu$m emission band, a satellite of the 3.3~$\mu$m aromatic C–H band \cite{Joblin1996}. Since aliphatic C–H bonds are more fragile than aromatic ones, this interpretation accounts for the observed variations in the 3.4/3.3~$\mu$m band intensity in star-forming regions exposed to significant UV irradiation \cite{Joblin1996, Pilleri2015, Peeters2024, Chown2024}. Chemical models generally predict that only large PAHs, typically containing 50 or more carbon atoms, can survive in regions where AIB emission is observed \cite{Allain1996,Montillaud2013,Andrews2016}. However, small PAHs have been detected in the gas phase of dark molecular clouds, and both small and methylated PAHs are present in the solid phase in meteorites. In molecular clouds and protoplanetary disks, including the protosolar disk, PAHs likely cycle between the gas and solid phases\cite{Derenne2005, Lange2023, Wenzel2024b}. Wenzel {\it{et al.}} \cite{Wenzel2024b} have modeled pyrene depletion on cold grains following its gas-phase formation but did not consider possible desorption mechanisms. In protoplanetary disks, Lange {\it{et al.}} \cite{Lange2023} employed a simple statistical approach to model PAH desorption into the gas phase via UV irradiation in the disk photosphere. In dark molecular clouds such as TMC-1, cosmic rays drive ionization. Here, desorption could occur through direct cosmic-ray interactions or be induced by secondary UV photons \cite{Shen2004, Dartois2023}.

The interaction of PAHs with UV photons in PDRs can trigger various processes, including ionization, dissociation, and radiative cooling \cite{Joblin2020}. The rates of these processes have been incorporated into models based on laboratory experiments and statistical approaches \cite{Allain1996,Montillaud2013,Andrews2016}. However, these models have yet to account for isomerization processes, which quantum chemical calculations suggest may play a key role in the photophysics of interstellar PAHs \cite{Parneix2017, Trinquier2017}. The recent discovery of small PAHs in dark clouds has further renewed interest in isomerization processes, particularly in the context of dissociative ionization of small PAHs and nitrogen-containing PANHs \cite{Arun2023,Banhatti2022,Bouwman2016}. Experimentally studying isomerization induced by light is highly relevant to these astrophysical environments, though it remains technically challenging. Recently, \citet{Jacovella2023} investigated the interconversion between azulene and naphthalene ions (\ch{C10H8+}). By probing chemical reactivity, they were able to reveal the presence of additional isomers forming when \ch{C10H8+} is exposed to 6 eV photons.

\begin{figure}[ht]
    \centering
   \includegraphics[width=0.7\linewidth]{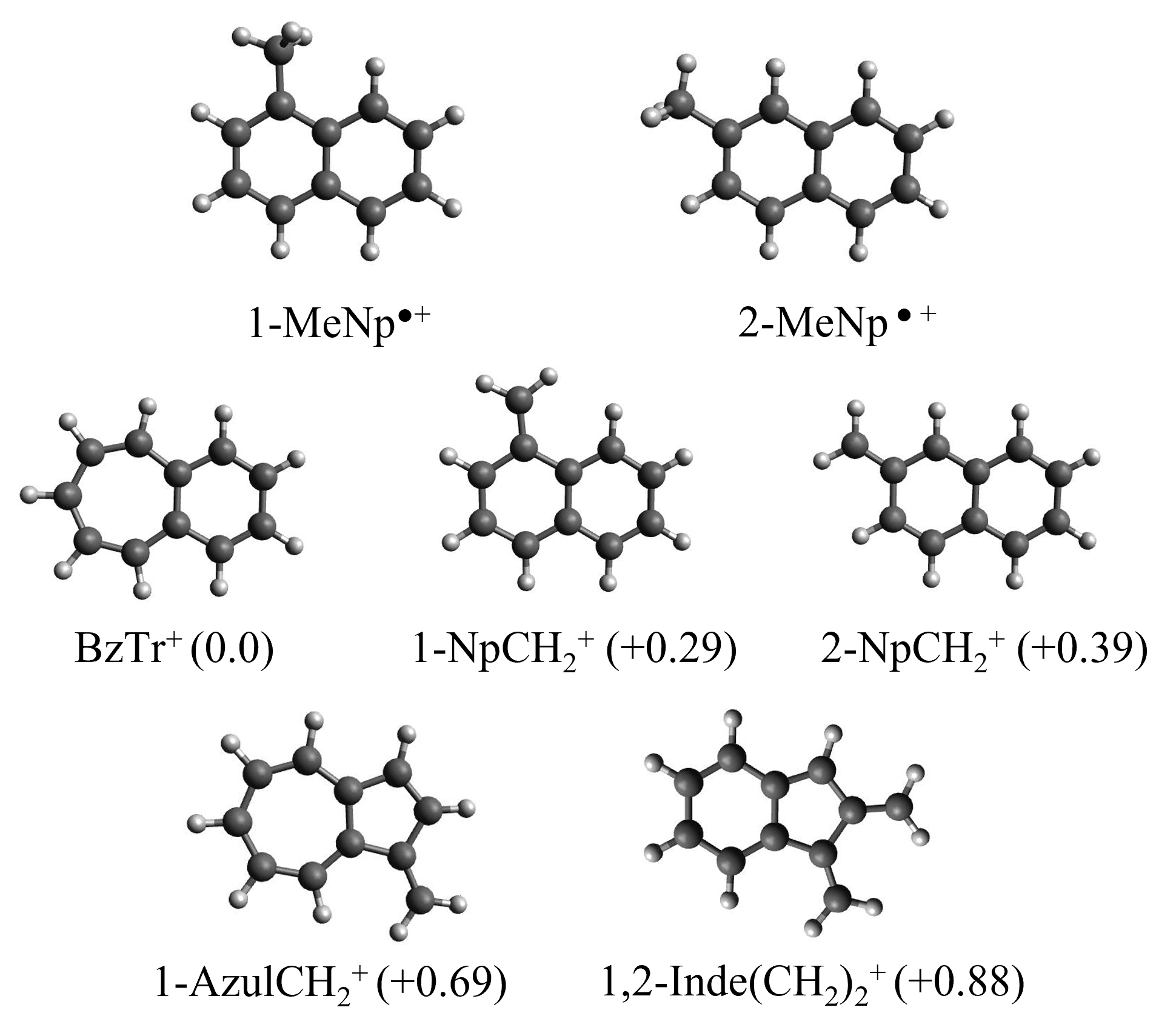}
    \caption{Top row: molecular structures of the 1- and 2-methylnaphthalene cations (1- and 2-\ch{MeNp^{.}+}). Middle and bottom rows: molecular structures of the five expected lowest-energy isomers resulting from H loss. The numbers within parenthesis correspond to relative energies, in eV, with respect to the lowest energy isomer, \ch{BzTr+}. The geometries were optimized at the B3LYP/6-31G(d,p) level of theory.}
    \label{geoms}
\end{figure}

The interaction of VUV photons with methylated PAHs results in the formation of a cationic -H fragment \cite{Gotkis1993,Jochims1999,Marciniak2021}. Gotkis {\it{et al.}} employed time-resolved photoionization mass spectrometry combined with ion-molecule reactions to investigate the \ch{C11H9+} isomers produced from 1- and 2-methylnaphthalene (1- and 2-MeNp) \cite{Gotkis1993}. Their study concluded that at least 20\% of the fragments are naphthylmethylium (\ch{NpCH2+}) and the rest benzyltropylium (\ch{BzTr+}) isomers. These isomers are considered to be the expected lowest-energy structures (see Figure~\ref{geoms}), analogous to the well-known benzylium and tropylium forms of \ch{C7H7+} generated from toluene \cite{Dunbar1975, Jusko2018, Torma2019, Rossi2023}. 

\citet{Wenzel2022} used infrared spectroscopy to characterize long-lived \ch{C11H9+} isomers derived from 2-MeNp (\ch{C10H7-CH3}) and 2-methylanthracene (\ch{C14H9-CH3}) \cite{Wenzel2022}. Their analysis identified two distinct species: 2-naphthylmethylium (2-\ch{NpCH2+}) and benzyltropylium (\ch{BzTr+}), the latter featuring a seven-membered ring (see Figure~\ref{geoms}). Similarly, \citet{Nagy2011} recorded the electronic spectrum of \ch{C11H9+} produced from 1-MeNp in Ne matrices, revealing clear vibronic structures attributed to 2-\ch{NpCH2+} and \ch{BzTr+} \cite{Nagy2011}. Notably, spectral signatures of 1-\ch{NpCH2+} were absent, despite theoretical predictions indicating strong transitions within the studied range.

In this work, we present a combined experimental and theoretical study to gain insights into the populations of long-lived \ch{C11H9+} isomers produced from 1-MeNp in the gas phase. Our approach integrates photodissociation spectroscopy with kinetic measurements of both photoprocessing and ion-molecule reactions. This study was made possible by leveraging the multiplex capabilities of PIRENEA, a setup dedicated to laboratory astrophysics. In addition to characterizing the spectral features of the different isomers, we also obtained insights into their photophysical behavior and provided evidence for interconversion between isomers occurring below the dissociation threshold. Finally, we discuss the broader perspectives and astrophysical implications of our findings.

%%%%%%%%%%%%%%%%%%%%%%%%%%%%%%%%%%%%%%%%%%%%%%%%%%%%%%%%%%%%%%%%%%%%%
%% Experimental and theoretical methods
%%%%%%%%%%%%%%%%%%%%%%%%%%%%%%%%%%%%%%%%%%%%%%%%%%%%%%%%%%%%%%%%%%%%%

\section{Methods}
\label{Methods}

%%%%%%%%Experiments%%%%%%%%%%
\subsection{Experimental methods}
\label{Experiment}

The experiments were performed using the PIRENEA setup, which has been described in detail elsewhere\cite{Useli-Bacchitta2010,Marciniak2021,Vinitha2022}. 
PIRENEA is a home-built Fourier Transform Ion Cyclotron Resonance mass spectrometer (FTICR-MS), which has been designed to study the photophysics and chemistry of PAHs in conditions relevant for the interstellar medium, in particular isolation on long timescales and low temperature \cite{Joblin2002}.

\begin{figure}[ht]
    \centering
    \includegraphics[width=1\linewidth]{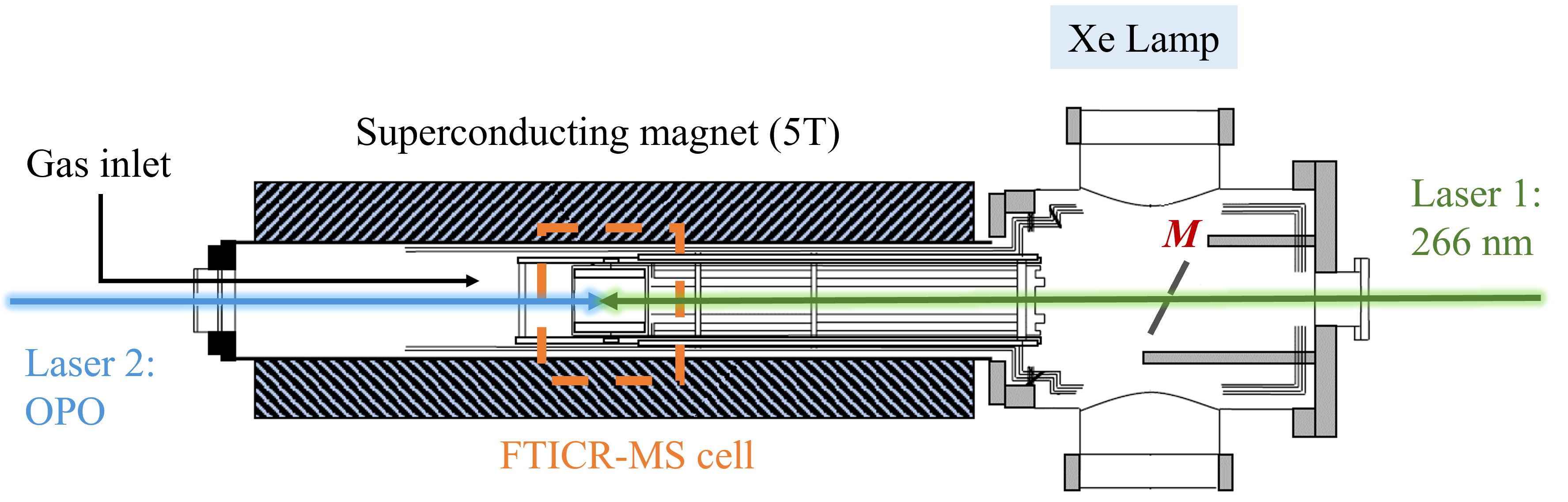}
    \caption{Scheme of the FTICR-MS PIRENEA setup. The \ch{C11H9+} ions are generated from a 1-MeNp vapor by irradiation at 266~nm (Laser 1). Laser 2, an optical parametric oscillator (OPO), is employed in the visible range for multiple-photon dissociation spectroscopy and photoprocessing. The parabolic mirror, denoted as $M$, focuses the Xe lamp irradiation into the ICR cell. For further details, see the text. }
    \label{PIRENEA}
\end{figure}

In the present experiments, we produce \ch{C11H9+} ions by using 1-MeNp as the precursor (Figure~\ref{geoms}). It has a high vapor pressure at room temperature but is highly condensable at cryogenic temperatures. The PIRENEA setup was therefore used at room temperature to avoid drastic condensation of the molecules on the walls. A vapor of the precursor was admitted into the experimental vacuum chamber. Irradiation with a 266~nm (4.66~eV) Nd:YAG Surelite I laser (Continuum; laser 1 in Figure~\ref{PIRENEA}) allowed the production of ions, particularly within the ICR cell where they were trapped. A typical mass spectrum after irradiation at 266~nm is shown in Figure~S1.
The absorption of at least two photons is required to ionize 1-MeNp (IP = 7.96~eV\cite{Jin2020, Gotkis1993, NISTChemistryWebBook}).  Jochims {\it{et al.}} \cite{Jochims1999} studied the photostability of methylated PAHs and derived a dissociation energy of 3.9~eV for 1-\ch{MeNp^{.}+} and 2-\ch{MeNp^{.}+} at an astrophysically critical dissociation rate of $10^2$~s$^{-1}$, which is relevant for the conditions in our experiments.  Therefore, the production of the -H fragment at \textit{m/z} 141 from the precursor 1-MeNp requires the absorption of at least 3 photons at 266~nm. This can happen within the same laser pulse or in a subsequent pulse after ionization. In some experiments, we used a Xe lamp with different filters as a gentle method to dissociate \ch{C11H10^{.}+} and generate \ch{C11H9+} ions. This approach also minimizes additional processing by 266 nm irradiation, which predominantly leads to the photodissociation of \ch{C11H9+}. 

Afterward, selective excitation in cyclotron motion was used to isolate the species with \textit{m/z} 141 by ejection of the other species trapped. 
The isolated \ch{C11H9+} species were then irradiated with laser~2, a mid-tunable optical parametric oscillator (OPO) laser (Panther EX OPO from Continuum) running at a frequency of 10 Hz with 5~ns pulse duration (Figure~\ref{PIRENEA}). In some experiments, laser~1 was also used to photoprocess the ions.
The typical time between the production and excitation of \ch{C11H9+} species was 3 seconds. Indeed, the effect of this time on the photofragmentation yield was investigated (up to 180~s) and no significant differences were observed.
After laser irradiation, an FTICR mass spectrum was recorded and used to derive a fragmentation yield as a function of the irradiation conditions (wavelength, laser energy, time between pulses and number of pulses; see Table~\ref{tableFigures}).

Laser~2 is collimated in a typical spot size of 0.24~cm$^2$. The typical number of photons per cm$^2$ for a pulse of energy  \( E_{\rm laser}=5.5~\rm{mJ} \) is $5.5\times10^{16}$ at 480~nm, which corresponds to a photon flux of $1.1\times10^{25}$ for a pulse duration of 5~ns.

To perform multiple-photon dissociation (MPD) spectroscopy, the dissociation yield was measured as a function of wavelength for a fixed number of pulses and fixed energy \( E_{\rm laser} \). In addition, photoprocessing kinetic curves were obtained by recording the normalized parent ion intensity as a function of the number of pulses at a fixed wavelength and energy \( E_{\rm laser} \). 
To account for variations in the absolute number of ions from one measurement to the other, the intensity of each ion \( M_{i} \) was normalized to the summed ion intensity following:

\begin{equation}
  I^{norm}_{M_i} =\frac{I_{M_{i}}}{\sum_{M_j}I_{M_j}} 
  \label{eq1}
\end{equation}

where \( I_{M_{i}} \) represents the intensity of the \textit{m/z} \( M_{i} \) and \( {\sum_{M_j}I_{M_j}} \)  represents the sum of all \textit{m/z} present (parent plus fragments). We note, however, that in most cases the conditions were very stable, allowing the use of absolute ion intensities as well (see Figure~S4 as an example). As previously described \cite{Useli-Bacchitta2010}, the most critical aspects concern the overlap between the laser beam and the ion cloud inside the ICR cell, as well as the laser beam profile that can affect the measured dissociation yield.  (see ref.\cite{Useli-Bacchitta2010} for further details). For this work, we used an improved version of the OPO beam alignment with automatic tracking to correct for changes in position with wavelength. This new feature limits the error associated with changes in the overlap of the laser beam with the ion cloud.

Some of the experiments presented here involve ion-molecule reactions with the neutral precursor (1-MeNp) to differentiate non-reactive isomers from reactive isomers. 
Due to the sticky nature of 1-MeNp, we are unable to report a precise value for the reactant pressure or fully control it. Typical pressure used in reactivity measurements was 10$^{-8}$~mbar.

%%%%%%%%Computational%%%%%%%%%%
\subsection{Computational approaches}
\label{theory}

The dissociation dynamics of the isomers of interest, \ch{C11H10^{.}+} and \ch{C11H9+}, were studied by molecular dynamics (MD) simulations using the methodology developed by Simon \textit{et al.} \cite{Simon2017} to describe the dissociation of PAHs in their ground electronic state  with initial internal energy under the form of vibrational energy. This amount of energy must be sufficient for dissociation to be observed within the nanosecond timescale.
This approach has been used to rationalize experimental results for a number of PAH-related species \cite{Jusko2018b,Rapacioli2018,Simon2018,Banhatti2022,Rap2023}. 
Briefly, the MD simulations are run in the Born-Oppenheimer approximation with the electronic structure described on-the-fly within the self consistent charge density functional based tight binding (SCC-DFTB) framework \cite{Elstner1998} 
using the modified set of C-H parameters recently published. \cite{Joblin2020}
The explicit description of the electronic structure allows the description of chemical evolution such as isomerization and dissociation reactions.
This approach is simply referred to as MD/DFTB in the rest of the manuscript.

In the present work, 720 simulations of 1\,ns with internal energies of 13 and 14\,eV were run for 1- and 2-\ch{MeNp^{.}+}. The values of internal energies were chosen as lowest energy values to obtain a reasonable quantitative fragmentation within the nanosecond timescale. Regarding the dissociation of \ch{C11H9+}, 180 simulations of 1\,ns with internal energies of 12.5 down to 11.6\,eV were run for 1-, 2-\ch{NpCH2+} and \ch{BzTr+} so as to obtain only one isomer resulting from the loss of \ch{C2H2}. Each simulation is performed from the ground-state geometry with random initial velocities in the microcanonical (NVE) ensemble.  Energies and gradients are computed every 0.1\,fs. This allows us to gain insights into reaction kinetics, branching ratios, and mechanisms. The results are described in section \ref{MD}. All MD/DFTB simulations were performed using the deMonNano code.\cite{deMon}

%%%%%%%%%%%%%%%%%%%%%%%%%%%%%%%%%%%%%%%%%%%%%%%%%%%%%%%%%%%%%%%%%%%%%
%% Results
%%%%%%%%%%%%%%%%%%%%%%%%%%%%%%%%%%%%%%%%%%%%%%%%%%%%%%%%%%%%%%%%%%%%%

\section{Results}
\label{Results}

While isomers cannot be distinguished by mass spectrometry, their differing absorption properties as a function of wavelength can be used to identify them. To gain insights into the long-lived isomers of dehydrogenated 1-MeNp cations (\ch{C11H9+}), we conducted a series of experiments complemented by MD/DFTB simulations. In particular, we focused on the spectral range of 430-590~nm (2.10-2.88~eV), where the lowest-energy isomers of \ch{C11H9+} are expected to absorb \cite{Nagy2011}. This range was investigated experimentally through a combination of multiphoton dissociation (MPD) spectroscopy, dissociation kinetics, and ion-molecule reactivity studies. Given the variety of experimental parameters used in our study, we have compiled the key details in Table~\ref{tableFigures} for clarity and reference.

The results are organized into three sections. First, we present a spectral analysis of the long-lived \ch{C11H9+} species. The second section focuses on quantifying the relative abundances of each isomer using reactivity measurements and photodissociation kinetics. Furthermore, we examine evidence for interconversion between isomers. Finally, we explore MD/DFTB simulations to gain insights into the dissociation dynamics of both MeNp radical cations (\ch{C11H10^{.}+}) and \ch{C11H9+} fragments.

\begin{table*}[!ht]
    \centering
    \begin{adjustbox}{width=20cm,  center}
    \begin{tabular}{!{\vrule width 2pt}c||c|c|c|c|c|c||c|p{4.5cm}||c|c!{\vrule width 2pt}}
    \noalign{\hrule height 2pt}
     \multicolumn{1}{ !{\vrule width 2pt}c|| }{\textbf{Figure}} & \multicolumn{6}{c ||}{\textbf{Production {\it{m/z}} 141}} & \multicolumn{2}{c||}{\textbf{OPO irradiation}} & \multicolumn{2}{c!{\vrule width 2pt}}{\textbf{Reactivity}} \\
    \noalign{\hrule height 2pt}
    \multirow{4}{1.5cm}{Fig.~\ref{MPDspectrum} (purple dots)} & \multicolumn{2}{c|}{\multirow{2}{*}{1-MeNp + 266~nm}} & \multicolumn{2}{c|}{\multirow{4}{*}{}} & \multicolumn{2}{c||}{\multirow{4}{*}{}} & $\lambda$ & 430-590~nm (2~nm steps) & \multicolumn{2}{c!{\vrule width 2pt}}{\multirow{4}{*}{}} \\
    \cline{8-9}
    & \multicolumn{1}{c}{}& & \multicolumn{1}{c}{} & & \multicolumn{1}{c}{} & & pulses & 20 & \multicolumn{1}{c}{} & \\
    \cline{2-3} \cline{8-9}
    & \( t_{\rm irr}\)& 24~s & \multicolumn{1}{c}{} & & \multicolumn{1}{c}{} & & delay & 100~ms & \multicolumn{1}{c}{} & \\
    \cline{2-3} \cline{8-9}
    & \( E_{\rm laser} \)& 4.5~mJ & \multicolumn{1}{c}{}& &\multicolumn{1}{c}{} & & \( E_{\rm laser} \) & 5.5~mJ & \multicolumn{1}{c}{} & \\
        \noalign{\hrule height 2pt}
     \multirow{5}{1.5cm}{Fig.~\ref{MPDspectrum} (orange stars)} & \multicolumn{2}{c|}{\multirow{2}{*}{1-MeNp + 266~nm}} & \multicolumn{2}{c|}{\multirow{4}{*}{}} & \multicolumn{2}{c||}{\multirow{4}{*}{}} & $\lambda$ & 430-550~nm (2~nm steps)  & \multicolumn{2}{c!{\vrule width 2pt}}{\multirow{4}{*}{}} \\
    \cline{8-9}
    & \multicolumn{1}{c}{}& & \multicolumn{1}{c}{} & & \multicolumn{1}{c}{} & & pulses & 5 & \multicolumn{1}{c}{} & \\
    \cline{2-3} \cline{8-9}
    & \( t_{\rm irr}\)& 24~s & \multicolumn{1}{c}{} & & \multicolumn{1}{c}{} & & delay & 2~s & \multicolumn{1}{c}{} & \\
    \cline{2-3} \cline{8-9}
    & \( E_{\rm laser} \)& 4.5~mJ & \multicolumn{1}{c}{}& &\multicolumn{1}{c}{} & & \( E_{\rm laser} \) & 5.5~mJ & \multicolumn{1}{c}{} & \\
        \noalign{\hrule height 2pt}
     \multirow{5}{1.5cm}{Fig.~\ref{spectraisomers} (a)} & \multicolumn{2}{c|}{\multirow{2}{*}{1-MeNp + 266~nm}} & \multicolumn{2}{c|}{\multirow{2}{*}{Xe lamp}} & \multicolumn{2}{c||}{\multirow{3}{*}{{\it m/z} 141 + 1-MeNp}} & $\lambda$ & 430-490~nm (2~nm steps) & \multicolumn{2}{c!{\vrule width 2pt}}{\multirow{5}{*}{}} \\
    \cline{8-9}
    & \multicolumn{1}{c}{}& & \multicolumn{1}{c}{} & & \multicolumn{1}{c}{} & & pulses & 600 & \multicolumn{1}{c}{} & \\
    \cline{2-5} \cline{8-9}
    & \( t_{\rm irr}\)& 6~s & \( t_{\rm irr}\) & 20~s & \multicolumn{1}{c}{} & & delay & 100~ms & \multicolumn{1}{c}{} & \\
    \cline{2-9} 
    & \( E_{\rm laser} \)& 4.5~mJ & Filters & CF$^a$ + ODF* & t & 30~s & \( E_{\rm laser} \) & 5.5~mJ & \multicolumn{1}{c}{} &  \\
        \noalign{\hrule height 2pt}
     \multirow{5}{1.5cm}{Fig.~\ref{spectraisomers} (b)} & \multicolumn{2}{c|}{\multirow{2}{*}{1-MeNp + 266~nm}} & \multicolumn{2}{c|}{\multirow{2}{*}{Xe lamp}} & \multicolumn{2}{c||}{\multirow{2}{*}{{\it m/z} 141 + 266~nm}} & $\lambda$ & 534-586~nm (2~nm steps)& \multicolumn{2}{c!{\vrule width 2pt}}{\multirow{5}{*}{}} \\
    \cline{8-9}
    & \multicolumn{1}{c}{}& & \multicolumn{1}{c}{} & & \multicolumn{1}{c}{} & & pulses & 55 & \multicolumn{1}{c}{} & \\
    \cline{2-9} 
    & \( t_{\rm irr}\)& 6~s & \( t_{\rm irr}\) & 20~s & \( t_{\rm irr}\) &  50~s & delay & 2~s & \multicolumn{1}{c}{} & \\
    \cline{2-9} 
    & \( E_{\rm laser} \)& 4.5~mJ & Filters & CF$^b$ & \( E_{\rm laser} \) & 4.5~mJ & \( E_{\rm laser} \) & 7~mJ & \multicolumn{1}{c}{} &  \\
            \noalign{\hrule height 2pt}

        \multirow{5}{1.5cm}{Fig.~\ref{reactivitycurve}} & \multicolumn{2}{c|}{\multirow{2}{*}{1-MeNp + 266~nm}} & \multicolumn{2}{c|}{\multirow{4}{*}{}} & \multicolumn{2}{c||}{\multirow{4}{*}{}} & \multicolumn{2}{c||}{\multirow{4}{*}{}}  & \multicolumn{2}{c!{\vrule width 2pt}}{\multirow{3}{*}{{\it m/z} 141 + 1-MeNp}} \\
    & \multicolumn{1}{c}{}& & \multicolumn{1}{c}{} & & \multicolumn{1}{c}{} & & \multicolumn{1}{c}{} & \multicolumn{1}{c||}{} & \multicolumn{1}{c}{} & \\
    \cline{2-3} 
    & \( t_{\rm irr}\)& 24~s & \multicolumn{1}{c}{} & & \multicolumn{1}{c}{} & & \multicolumn{1}{c}{} & \multicolumn{1}{c||}{} & \multicolumn{1}{c}{} & \\
    \cline{2-3} \cline{10-11}
    & \( E_{\rm laser} \)& 4.5~mJ & \multicolumn{1}{c}{}& &\multicolumn{1}{c}{} & & \multicolumn{1}{c}{} & \multicolumn{1}{c||}{} & t & 1-240~s \\
        \noalign{\hrule height 2pt}
     \multirow{5}{1.5cm}{Fig.~\ref{PCK1MeNp}} & \multicolumn{2}{c|}{\multirow{2}{*}{1-MeNp + 266~nm}} & \multicolumn{2}{c|}{\multirow{4}{*}{}} & \multicolumn{2}{c||}{\multirow{4}{*}{}} & $\lambda$ & 444, 454, 480, 516, 538, 548 and 564~nm & \multicolumn{2}{c!{\vrule width 2pt}}{\multirow{4}{*}{}} \\
    \cline{8-9}
    & \multicolumn{1}{c}{}& & \multicolumn{1}{c}{} & & \multicolumn{1}{c}{} & & pulses & 0-10000 & \multicolumn{1}{c}{} & \\
    \cline{2-3} \cline{8-9}
    & \( t_{\rm irr}\)& 24~s & \multicolumn{1}{c}{} & & \multicolumn{1}{c}{} & & delay & 100~ms & \multicolumn{1}{c}{} & \\
    \cline{2-3} \cline{8-9}
    & \( E_{\rm laser} \)& 4.5~mJ & \multicolumn{1}{c}{}& &\multicolumn{1}{c}{} & & \( E_{\rm laser} \) & 5.5~mJ & \multicolumn{1}{c}{} & \\
        \noalign{\hrule height 2pt}
     \multirow{5}{1.5cm}{Fig.~\ref{reactivityKinetic_log}} & \multicolumn{2}{c|}{\multirow{2}{*}{1-MeNp + 266~nm}} & \multicolumn{2}{c|}{\multirow{4}{*}{}} & \multicolumn{2}{c||}{\multirow{4}{*}{}} & $\lambda$ & 444, 454, 480, 516, and 564~nm  & \multicolumn{2}{c!{\vrule width 2pt}}{\multirow{3}{*}{{\it m/z} 141 + 1-MeNp}} \\
    \cline{8-9}
    & \multicolumn{1}{c}{}& & \multicolumn{1}{c}{} & & \multicolumn{1}{c}{} & & pulses & 40-6000 & \multicolumn{1}{c}{} & \\
    \cline{2-3} \cline{8-9}
    & \( t_{\rm irr}\)& 24~s & \multicolumn{1}{c}{} & & \multicolumn{1}{c}{} & & delay & 100~ms & \multicolumn{1}{c}{} & \\
    \cline{2-3} \cline{8-11}
    & \( E_{\rm laser} \)& 4.5~mJ & \multicolumn{1}{c}{}& &\multicolumn{1}{c}{} & & \( E_{\rm laser} \) & 5.5~mJ & t & 240~s \\
        \noalign{\hrule height 2pt}
    \multicolumn{11}{l}{$^a$Color Filter at 495~nm, $^b$Color Filter at 610~nm, *Optical Density Filter (37\% transmission).}\\
    \end{tabular}
    \end{adjustbox}
    \caption{Experimental conditions corresponding to each Figure presented in Sections~\ref{MPDspectroscopy}, \ref{PKCreactivity}, and \ref{Interconversion}.}
    \label{tableFigures}
\end{table*}

%%%%%%%%%%%%%%%%%%%%%%%%%%%%%%%%%%%%%%%%%%%%%%%%%%%%%%%%%%%%%%%%%%%%
%% MPD spectroscopy
%%%%%%%%%%%%%%%%%%%%%%%%%%%%%%%%%%%%%%%%%%%%%%%%%%%%%%%%%%%%%%%%%%%%

\subsection{Isomer differentiation using MPD spectroscopy}
\label{MPDspectroscopy}

MPD spectroscopy was performed by recording the dissociation yield of \textit{m/z} 141 as a function of the OPO wavelength. A single primary fragment at \textit{m/z} 115 (\ch{C9H7+}) was observed across all wavelengths. It results from the loss of neutral acetylene (\ch{C2H2}) from \ch{C11H9+}. This exclusive fragmentation pathway is consistent with previous studies \cite{Mordaun1984, Wenzel2022}. Below 470~nm, this fragment undergoes further dissociation (\ch{C2H2} loss), yielding a secondary fragment at \textit{m/z} 89 (see Figure~S3). Since indenyl cations are known to absorb in this wavelength range, we attribute the observed \ch{C9H7+} species to this cation. This assignment is further supported by our MD simulations, as discussed in section \ref{MD}.

\begin{figure}[!ht]
    \centering
    \includegraphics[width=0.6\linewidth]{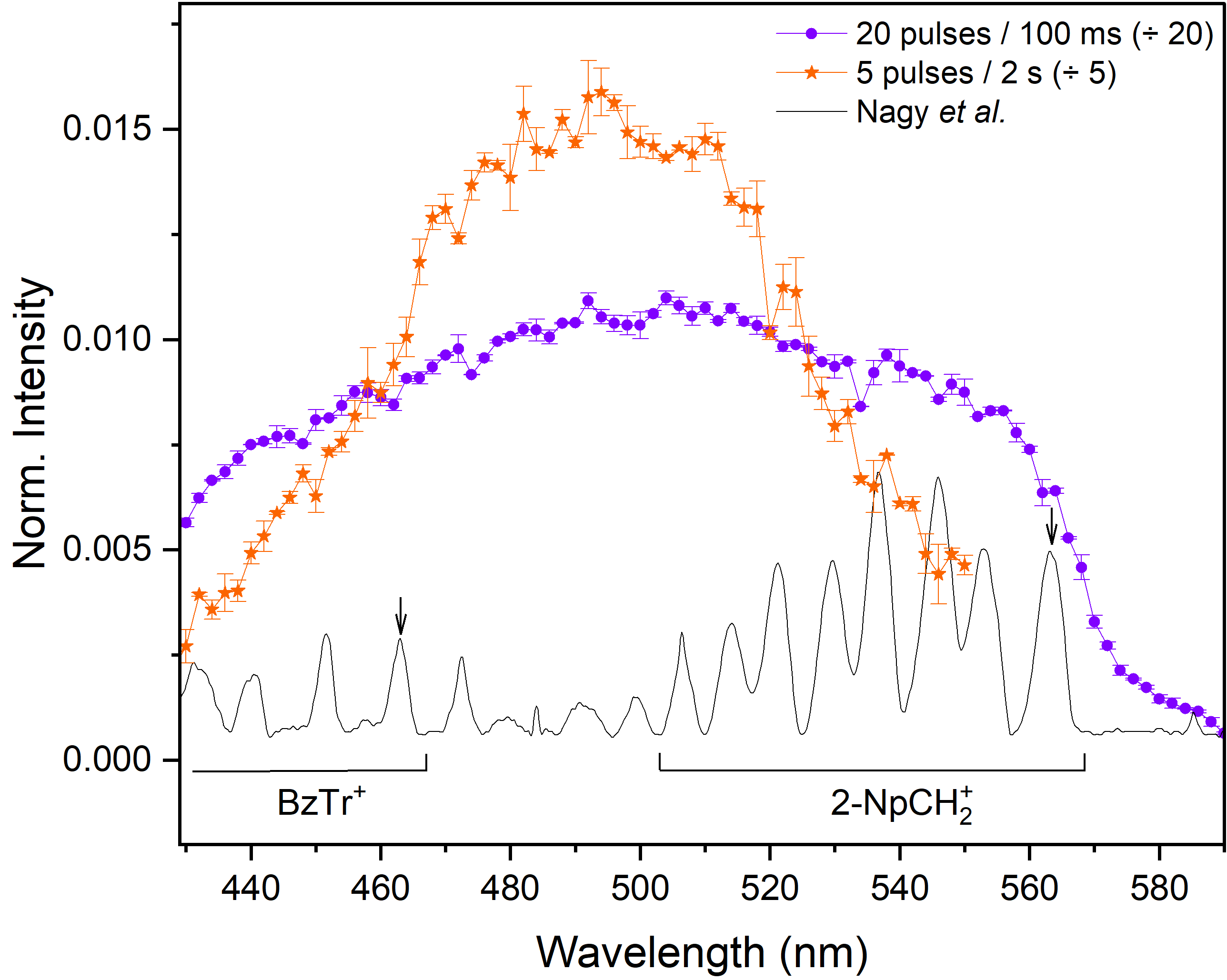}
    \caption{Measured MPD spectra of gas-phase \ch{C11H9+} produced from the 1-MeNp precursor. The spectra (purple dots and orange stars) correspond to 
    two different OPO irradiation conditions (Table~\ref{tableFigures}). For comparison purposes, the normalized intensity has been divided by the number of laser pulses. The photo-absorption spectrum of \ch{C11H9+} ions in Ne matrices at 6~K \cite{Nagy2011} is also represented in arbitrary units (black line). The black arrows indicate the band origins assigned by Nagy {\it et al.}\cite{Nagy2011}}.
    \label{MPDspectrum}
\end{figure}

Figure~\ref{MPDspectrum} shows the MPD spectra of \ch{C11H9+} recorded for two different OPO irradiation conditions (see Table~\ref{tableFigures}). Overall, the spectra exhibit a broad feature whose wavelength extent depends on the irradiation conditions. Each measurement was repeated twice showing no significant variations (as indicated by the error bars), which gives confidence in the observed structures. Additionally, the total ion intensity remained stable throughout the experiments (see Figure~S4).

The electronic absorption spectrum of the species \ch{C11H9+}, produced from 1-MeNp and isolated in a neon matrix at 6~K, is also included in Figure~\ref{MPDspectrum} \cite{Nagy2011}. Above 420~nm, two distinct bands were identified by the authors based on theoretical calculations: one between 420 and 480~nm, attributed to \ch{BzTr+}, and another above 500~nm, peaking near 540~nm, attributed to 2-\ch{NpCH2+}. While these bands are likely present in our MPD spectra, they appear obscured by a broad and intense feature centered around 490~nm—precisely where the neon matrix spectrum reported by \citet{Nagy2011} shows a minimum. This striking difference cannot be explained by temperature effects or a simple spectral shift. Instead, it suggests that at least one additional isomer is present in our experiments with strong absorption in the 460–530~nm range.

Therefore, we assume that our MPD spectra result from the superposition of more than two isomers. To support this assumption and gain a clearer understanding of the contribution of each isomer, in the following, we present measurements aimed at improving the comparison with the photo-absorption spectrum\cite{Nagy2011}.

\begin{figure}[H]
    \centering
    \includegraphics[width=1\linewidth]{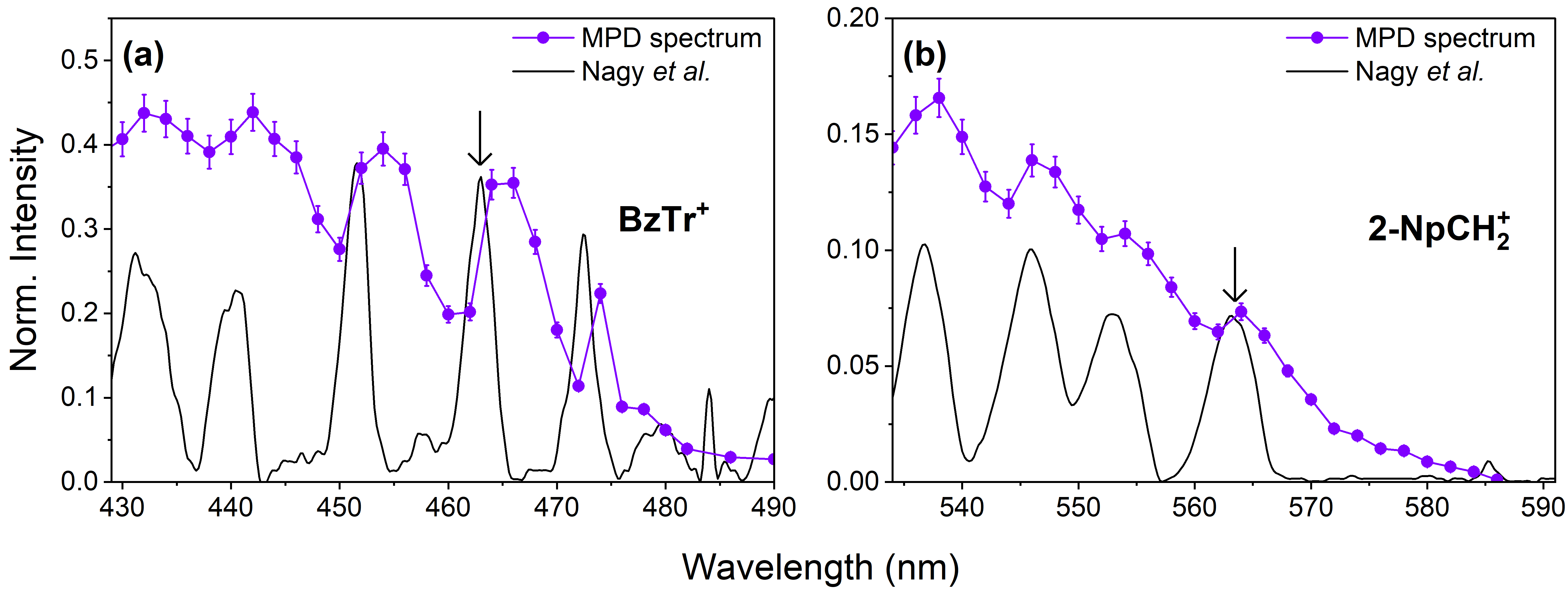}
    \caption{Measured MPD spectra of gas-phase \ch{C11H9+} ions (purple dots) produced from the 1-MeNp precursor (see Table~\ref{tableFigures}). The identification with the different isomers of \ch{C11H9+} is based on previous work \cite{Nagy2011}. The photo-absorption spectrum measured by \citet{Nagy2011} in Ne matrices is also represented (black line). The black arrows indicate the band origins derived by these authors (see Table~\ref{tablebandmaxima}). The observed band maxima of (a) \ch{BzTr+} and (b) 2-\ch{NpCH2+} are given in Table~\ref{tablebandmaxima}.}
    \label{spectraisomers}
\end{figure}

In a first set of experiments, we purified the isomer mixture by using ion-molecule reactions. The latter have been proved to be a powerful tool for distinguishing \ch{Bz+} from \ch{Tr+} among \ch{C7H7+} isomers \cite{Dunbar1975, Shen1974, Zins2010,Melko2013}. The reactivity of \ch{Bz+} with toluene leads to the formation of \ch{C8H9+}, whereas \ch{Tr+} is unreactive. A similar trend is expected in the case of \ch{C11H9+} isomers (see section~\ref{PKCreactivity}). Figure~\ref{spectraisomers} (a) shows the MPD spectrum obtained when allowing sufficient reaction time to ensure that most reactive ions have undergone reaction (see Table~\ref{tableFigures}). The MPD spectrum was measured using a long irradiation time (60~s at 10~Hz), with the goal of dissociating as many ions as possible. Under these conditions, we succeeded in dissociating approximately 50\% of the trapped ions. As described in previous work using the PIRENEA setup\cite{Useli-Bacchitta2010}, the overlap between the laser beam and the ion cloud is a critical factor, and this overlap typically decreases over time due to ion cloud expansion. The 50\% level of dissociation is therefore plausible given the conditions. Additionally, the MPD spectrum clearly shows the vibronic progression previously observed in the photo-absorption spectrum and attributed to \ch{BzTr+} (see also Table~\ref{tablebandmaxima}). Intriguingly, our MPD spectrum also exhibits a local maximum at $474 \pm 1$~nm. This feature may correspond to the 472.6~nm band observed in the matrix photoabsorption spectrum. However, \citet{Nagy2011} attributed that band to a radical species -- a possibility ruled out by our ion-trap experiment. Notably, \citet{Nagy2011} also measured the fluorescence of \ch{BzTr+} and identified the zero-phonon line at 463.8~nm. One possible explanation for the structure observed on the red side of the origin band is the interplay between photon-induced heating and radiative cooling via fluorescence, which directly influences the dissociation yield. A tentative interpretation is that absorption at 474~nm occurs through hot bands, but the fluorescence efficiency at this wavelength is lower than at shorter wavelengths.

Since \ch{BzTr+} is unreactive, it cannot be selectively removed from the isomer population. However, we found that its abundance can be minimized through photoprocessing at 266~nm (Table~S1). This approach was inspired by the findings of Nagy {\it et al.} \cite{Nagy2011}, which suggest that \ch{BzTr+} has the highest absorption cross-section at this wavelength. (see, for instance, Figure~9 in ref.\cite{Nagy2011}). This method allows us to reduce the \ch{BzTr+} population to $\sim$14~\% (see Table~S1).

\begin{table}[]
    \centering
     \begin{tabular}{|c||c|c|}
      \hline
        Isomer & $\lambda_{max}$~(nm) ($I_{MPD}$) & $\lambda_{max}$~(nm) (Abs.) \cite{Nagy2011} \\
      \hline
       \multirow{7}{2.5cm}{\ch{BzTr+}}& $\mathit{474 \pm 1}$$^{(a)}$ & \\
       &  & $\mathit{463.8}$$^{ZPL}$ \\
      & $466 \pm 1$ ($\geq 5.9 \times 10^{-4}$) & 462.6$^{(b)}$ ($\sim0.018$) \\
      & $454 \pm 1$ & 451.1   \\
      & $442 \pm 1$ ($7.3 \times 10^{-4}$) & 440.1 ($\sim0.014$)   \\
      & $432 \pm 1$ & 430.9   \\
      & $426 \pm 1$ & 421.8  \\
      \hline
      \multirow{5}{2.5cm}{2-\ch{NpCH2+}} &  & $\mathit{563.2}$$^{ZPL}$ \\
      & $564.5 \pm 1$ ($1.3 \times 10^{-3}$) & 562.2$^{(b)}$ ($\sim0.032$) \\
      & $554.5 \pm 1$ & 552.1   \\
      & $547 \pm 1$   & 546.0   \\
      & $538 \pm 1$ ($3 \times 10^{-3}$)  & 536.1 ($\sim0.046$) \\
      \hline
      \multirow{3}{2.5cm}{1-\ch{NpCH2+}}& $\mathit{538 \pm 1}$$^{(a)}$   &    \\
      & $530 \pm 2$ ($1.2 \times 10^{-2}$)  &    \\
      & $494 \pm 1$ ($2.3 \times 10^{-2}$)  &    \\
      \hline
     \end{tabular}
         \caption{Spectroscopic features in the visible range for \ch{BzTr+}, 2-\ch{NpCH2+}, and 1-\ch{NpCH2+}, based on our MPD spectra and the absorption and fluorescence data from \citet{Nagy2011}. Band maxima ($\lambda_{\text{max}}$) are listed along with their intensities: normalized MPD intensities per pulse ($I_{MPD}$; this work) and absorbance values (Abs.; \citet{Nagy2011}). Superscript $^{(b)}$ marks absorption band origins identified in a neon matrix\cite{Nagy2011}. Italicized values indicate: $^{(a)}$ local maxima in the MPD spectra not assigned to vibronic progressions, and $^{ZPL}$ zero-phonon line positions from \citet{Nagy2011}. }
     \label{tablebandmaxima}
 \end{table}

Figure~\ref{spectraisomers}(b) displays the MPD spectrum recorded at longer wavelengths (535–590~nm), following processing at 266~nm and using extended OPO irradiation (110~s with a 2~s delay between pulses). The resolved bands shown in Figure~\ref{spectraisomers}(b) are attributed to 2-\ch{NpCH2+}, in line with the assignment proposed by \citet{Nagy2011}. As previously illustrated in Figure~\ref{MPDspectrum}, using a shorter delay between laser pulses (i.e., 100~ms) results in a broad and intense feature that hinders clear observation of the well-defined bands visible in Figure~\ref{spectraisomers}(b) and listed in Table~\ref{tableFigures}. This effect is likely due to the build-up of hot ion populations when using short delays between pulses (see discussion in Section~\ref{DissIsom}). In contrast, a longer delay of 2~s allows the ions to cool more effectively before being irradiated again, which favors dissociation within a single pulse and leads to sharper dissociation bands.\cite{Stockett2019}.

Figures~\ref{MPDspectrum} and S5 highlight the complexity of the broad spectral feature and its sensitivity to the OPO irradiation conditions. First, the dissociation yield is found to vary with irradiation parameters, an aspect that will be discussed in Section~\ref{DissIsom}. Second, the MPD spectrum remains congested despite employing a 2~s delay between laser pulses or even using a single pulse to record the action spectrum (Figure~S5).
We also verified that this broad feature is not caused by photo-processing by 266~nm irradiation during ion production. To do so, we first generated parent ions at \textit{m/z} 142 using 266~nm irradiation, then isolated \textit{m/z} 142 before exposing it to Xe lamp irradiation filtered at 610~nm to gently produce \textit{m/z} 141. We subsequently recorded a portion of the MPD spectrum and confirmed that the production method had no significant effect on the broad feature. However, this approach does affect the relative abundance of \ch{BzTr+}, which increases to approximately 50\% (Table~S1). Using 266~nm irradiation in the subsequent step (see conditions in Table\ref{tableFigures}) minimizes the \ch{BzTr+} population.

Due to significant variations in the positions of local maxima across the MPD spectra (Figure~S5), it is challenging to identify a consistent series of bands that can confidently be assigned as vibronic transitions. Therefore, we report only two specific band positions in Table~\ref{tableFigures}. The first is a well-defined band at $494 \pm 1$~nm, which is consistently observed across all spectra. The second, at $530 \pm 2$~nm, lies on the blue side of a dip in the dissociation yield followed by another band centered at $538 \pm 1$~nm. The two features at about 530 and 538~nm appear robust. By analogy with the case of \ch{BzTr+}, we propose that the band at $530 \pm 2$~nm corresponds to the origin of the first electronic transition of the isomer responsible for the intense broad feature. The associated fluorescence, combined with hot-band absorption, would then give rise to the dip at $534 \pm 1$~nm and the local maximum observed at $538 \pm 1$~nm. Notably, this position coincides with a vibronic band of 2-\ch{NpCH2+}, a possible explanation for which will be discussed in Section~\ref{DissIsom}.

Vertical excitation energies computed by Nagy et al. \cite{Nagy2011} at the CC2 level suggest that 2-\ch{NpCH2+} is responsible for the band with an origin at 561~nm observed in Ne matrices. However, two later theoretical studies \cite{Li2023, Troy2012} reassigned this band to the 1-\ch{NpCH2+} isomer. Our experimental results, however, support the original assignment by Nagy et al. Specifically, the intense and broad feature seen in our MPD spectra is best attributed to the 1-\ch{NpCH2+} isomer, for which theoretical calculations \cite{Nagy2011} predict two absorption bands with oscillator strengths approximately twice those of 2-\ch{NpCH2+} in the visible range. In Table~S3, we propose plausible gas-phase band origin positions based on our MPD spectra and Time-Dependent Density Functional Theory (TD-DFT) calculations. The values for \ch{BzTr+} and 2-\ch{NpCH2+} are more confidently assigned, as their vibronic structures are well-resolved. In contrast, the assignment for 1-\ch{NpCH2+} remains tentative due to the broad and congested nature of its spectral feature.

%%%%%%%%%%%%%%%%%%%%%%%%%%%%%%%%%%%%%%%%%%%%%%%%%%%%%%%%%%%%%%%%%%%%%
%% Kinetic and Ion-molecule reaction
%%%%%%%%%%%%%%%%%%%%%%%%%%%%%%%%%%%%%%%%%%%%%%%%%%%%%%%%%%%%%%%%%%%%%

\subsection{Quantifying the isomer populations}
\label{PKCreactivity}

In the previous section, we concluded that the \ch{C11H9+} isomer population consists of, at least, the three lowest-energy isomers: 1-\ch{NpCH2+}, 2-\ch{NpCH2+}, and \ch{BzTr+}. {Here, we estimate the relative abundances of the initial isomer populations prior to OPO irradiation.
These results were obtained by combining the results from ion-molecule reactions with those of photoprocessing kinetic curves, as presented below. We also report experimental evidence of interconversion between isomers.

\subsubsection{Ion-molecule reactions}

In our experiments, two primary products resulting from the reaction of \ch{C11H9+} with 1-MeNp were observed as illustrated in Figure~S2: (i) \ch{C12H11+} at \textit{m/z} 155, as reported previously \cite{Gotkis1993}, and (ii) \ch{C22H19+} at \textit{m/z} 283, which corresponds to an adduct (not reported previously). Additionally, we observed a peak at \textit{m/z} 297, corresponding to \ch{C23H21+}, which is the adduct of \ch{C12H11+} (\textit{m/z} 155) with 1-MeNp, formed through secondary reactions.
The formation of adducts observed in the case of \ch{C7H7+} studies was attributed to stabilization by collisions with He buffer gas \cite{Zins2010} as these adducts were not present in ICR experiments. On the opposite, we found that the larger \ch{C11H9+} ions can form adducts with 1-MeNp in isolated conditions.
A peak at {\it m/z} 142 was also observed, but it is unlikely to correspond to a charge transfer reaction channel, based on the calculated ionization potentials (IPs) of the five lowest-energy isomers of \ch{C11H9+} (see Table~S2). While some contribution from the fragmentation of adducts during detection cannot be ruled out, the majority of this peak likely results from a residual in the isolation of {\it m/z} 141 (see, for instance, Figure~S2). As a result, we did not consider this peak in our analysis.

A representative reactivity kinetic curve of \ch{C11H9+} with 1-MeNp is shown in Figure~\ref{reactivitycurve}. The curve was fitted using a two-exponential decay model with an additional constant term, which corresponds to the asymptotic value. This value stabilizes at approximately 0.2 when enough time is allowed for all reactive ions to undergo reaction. As a result, the remaining ions represent the non-reactive species, specifically \ch{BzTr+}, with a relative abundance estimated to be around 20\% for the production method used. Additionally, the presence of two distinct decay rates suggests the existence of two reactive populations that react at different rates but with comparable abundances of approximately 35\% and 45\%. These two populations may correspond to the two reactive isomers, 1-\ch{NpCH2+} and 2-\ch{NpCH2+}, which we identified through spectroscopy. In this case, the two isomers exhibit kinetic reaction rates differing by an order of magnitude, the most reactive ion corresponding to the 35\% relative abundance.

\begin{figure}[H]
    \centering
    \includegraphics[width=0.7\linewidth]{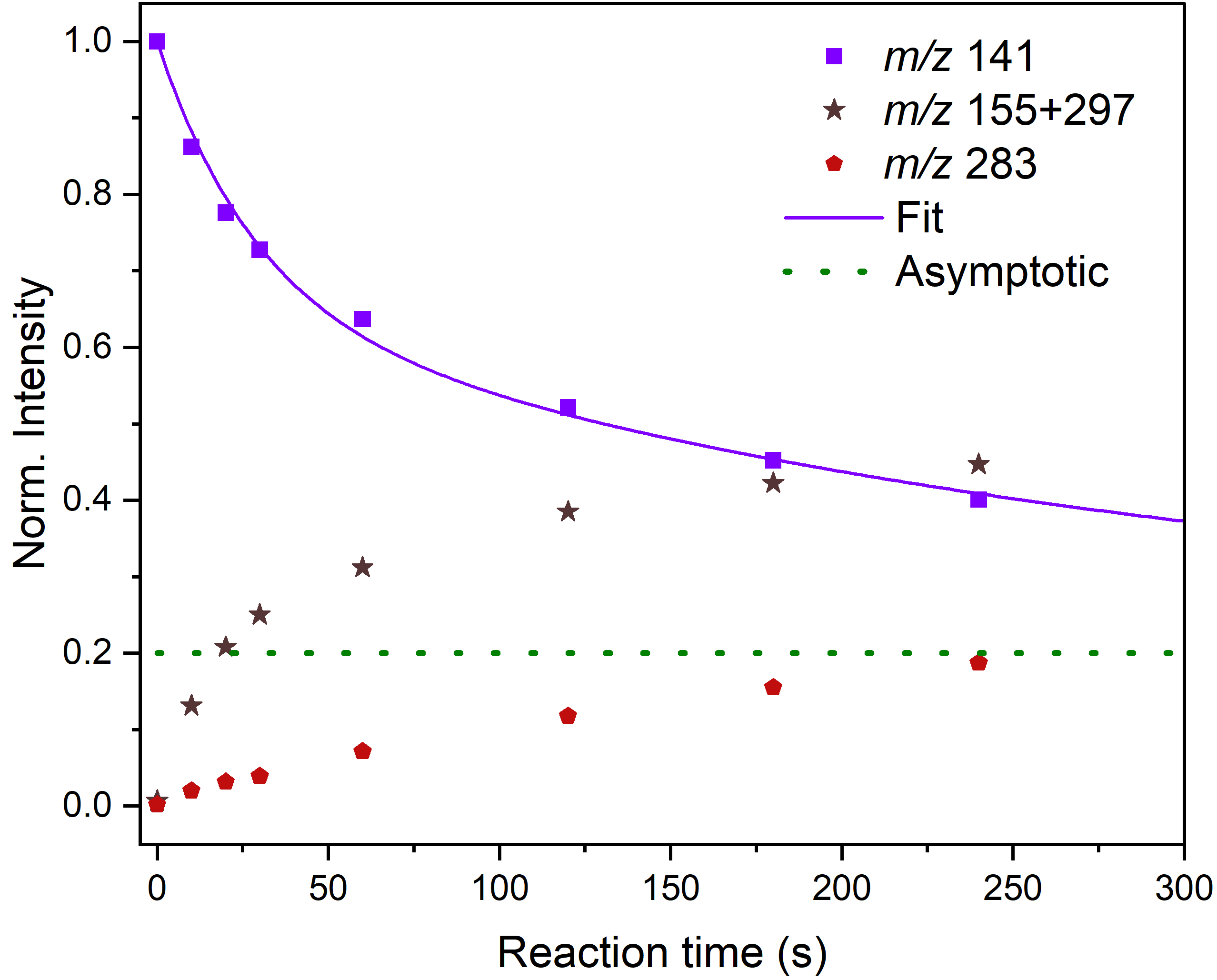}
    \caption{Reactivity kinetic curve of \ch{C11H9+} (purple squares) with 1-MeNp. The primary reaction channel leading to {\it m/z} 155, as well as the secondary channel at {\it m/z} 297 resulting from further reaction with the neutral precursor, are shown as brown stars. The red hexagons represent another primary reaction channel, which forms an adduct at {\it m/z} 283. The solid line represents a fit using a two-exponential decay model plus a constant, which corresponds to the experimental asymptotic limit (dotted line). The fitting parameters for the two-exponential decay are $A_{1} = 0.35$, $k_{1} = 0.035~s^{-1}$, and $A_{2} = 0.45$, $k_{2} = 0.0032~s^{-1}$.}
    \label{reactivitycurve}
\end{figure}

\subsubsection{Photoprocessing kinetic curves}
\label{Photoprocessing}

The three isomers absorb differently within the 430–590~nm range as presented in Section~\ref{MPDspectroscopy}. We therefore investigated the use of photofragmentation kinetics to gain insights into the isomer populations. The measurements were performed at a fixed laser energy \( E_{\rm laser} \) and at different wavelengths: 444, 454, 480, 516, 538, 548, and 564~nm. The photofragmentation kinetic curves (Figure~\ref{PCK1MeNp}) can be described by an initial rapid decay followed by a slower decay. At 1000~s of irradiation (10000 pulses), the normalized parent ion population has reached an asymptotic value for wavelengths above 500~nm, while it continues to decrease slightly for wavelengths below 500~nm. Furthermore, the value for the longest irradiation times decreases with decreasing wavelength, ranging from 0.46 at 564~nm to 0.16 at 444~nm. All kinetic curves were fitted using a three-exponential decay model, described by the system of differential equations~S1 in the SI.

The fitting parameters for the photofragmentation kinetic curves are presented in Table~\ref{tablePKC_1-MeNp}. As a first approximation, we attribute the fast decay component at 516 and 480~nm to the isomer responsible for the broad and intense dissociation band, which we identified as 1-\ch{NpCH2+}. Based on the corresponding $A_{1}$ values, we derive an average relative abundance of 37\% for 1-\ch{NpCH2+}.

\begin{figure}[!ht]
    \centering
    \includegraphics[width=1\linewidth]{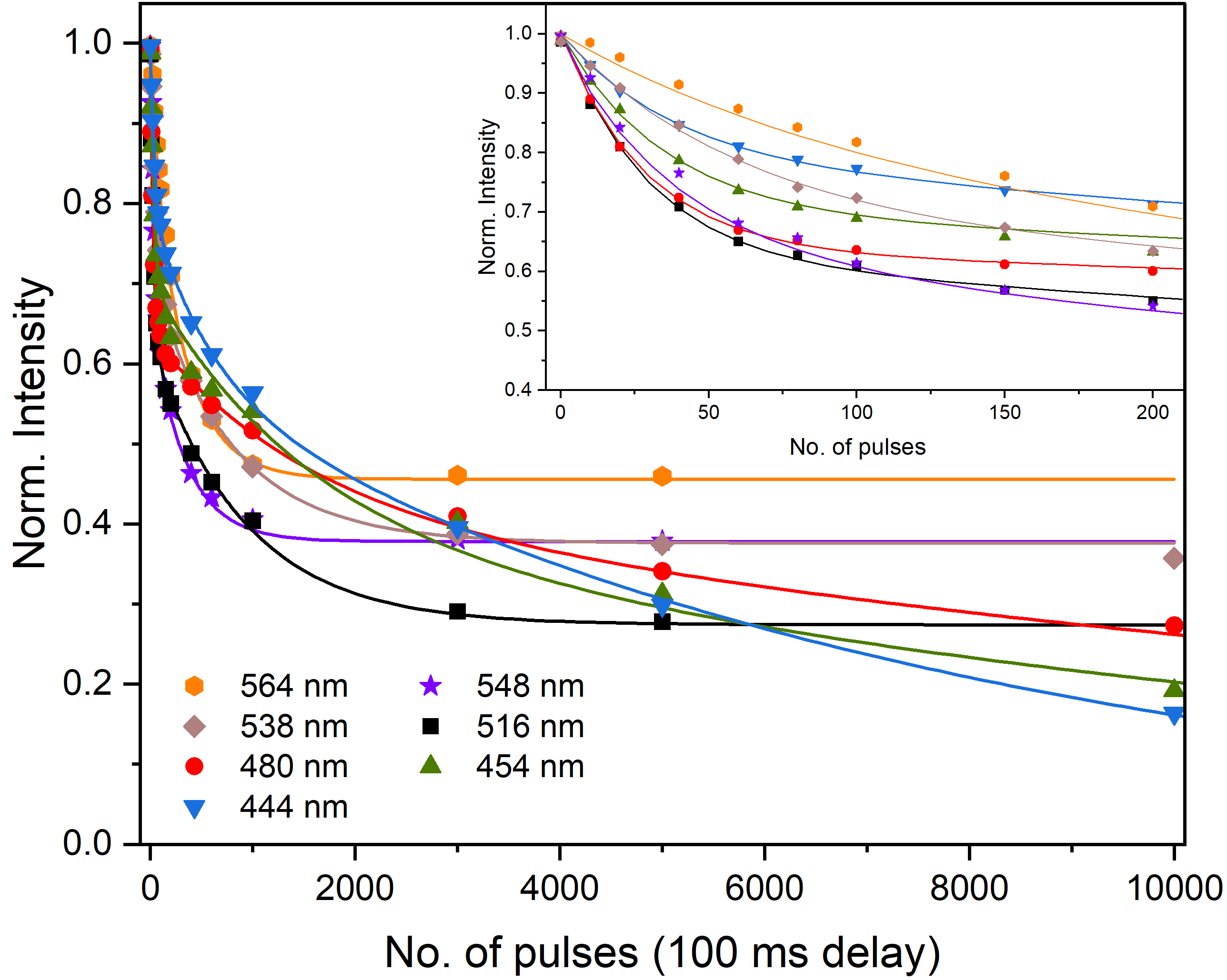}
    \caption{Photofragmentation kinetic curves of \ch{C11H9+} ions recorded at different wavelengths: 444~nm (blue inverted triangles), 454~nm (green triangles), 480~nm (red circles), 516~nm (black squares), 538~nm (light brown diamonds), 548~nm (purple stars), and 564~nm (orange hexagons). The \ch{C11H9+} ions were produced from 1-MeNp. The solid lines represent fits based on a three-exponential decay model, as described by Eq.~S1 (see SI for further details concerning the fitting procedure).}
    \label{PCK1MeNp}
\end{figure}

As we move away from the band maximum, both $k_{1}$ and $A_{1}$ decrease. While the drop in $k_{1}$ is expected due to a lower absorption cross-section, the significant reduction in $A_{1}$ is less intuitive if $A_{1}$ represents the initial population of 1-\ch{NpCH2+}. This behavior strongly suggests a competition between dissociation and isomerization processes for 1-\ch{NpCH2+}.

In the longest wavelength range (516~nm and above), 2-\ch{NpCH2+} shows absorption while \ch{BzTr+} does not. At 516~nm, the absorption cross-section of 2-\ch{NpCH2+} is likely lower than that of 1-\ch{NpCH2+}, making it a probable contributor to the $A_{2}$ and $k_{2}$ components. In the previous section, we tentatively assigned the origin band of 1-\ch{NpCH2+} at $530 \pm 2$~nm. Absorption at longer wavelengths (538, 548, 564 nm) indicates the presence of a hot 1-\ch{NpCH2+} population, likely arising from the isomerization of 2-\ch{NpCH2+}. Once formed, hot 1-\ch{NpCH2+} would absorb and dissociate efficiently. However, distinguishing the individual contributions of 1-\ch{NpCH2+} and 2-\ch{NpCH2+} to the $A_{1}$, $A_{2}$, $k_{1}$, and $k_{2}$ parameters remains challenging. Conversely, the $A_{3}$ component—representing the asymptotic value—can be confidently attributed to \ch{BzTr+}, whose initial relative abundance (approximately 0.2) was derived from reactivity measurements (see Figure~\ref{reactivitycurve}). The significantly larger $A_{3}$ values observed in the kinetics therefore provide strong evidence for isomerization of 1-\ch{NpCH2+} and/or 2-\ch{NpCH2+} into \ch{BzTr+}.

Applying a similar reasoning at shorter wavelengths (454 and 444 nm), the $A_{1}$, $A_{2}$, $k_{1}$, and $k_{2}$ parameters can be attributed to 1-\ch{NpCH2+} and \ch{BzTr+}, while $A_{3}$ is likely associated with 2-\ch{NpCH2+}. The observed variation in $A_{3}$ can then be attributed to the isomerization of 1-\ch{NpCH2+}. This assignment is supported by the trend that $A_{3}$ increases as the absorption by 1-\ch{NpCH2+} decreases, suggesting that under these conditions, isomerization is favored over direct dissociation. Additionally, the very low $k_{3}$ values indicate that 2-\ch{NpCH2+} has a weak absorption cross-section in this spectral region.

Finally, the photofragmentation kinetic curve of \ch{C11H9+} was also studied under laser irradiation at 266~nm (see Section \ref{Experiment} for further details). The resulting kinetic curve, shown in Figure~S7, exhibits distinct behavior compared to the visible range. At 266~nm, the entire ion population dissociates within 300~s of irradiation (3000 pulses), following a single-exponential decay, suggesting that all isomers dissociate with similar efficiency. However, this contradicts the observed photoprocessing at 266~nm, which results in a relative depletion of the \ch{BzTr+} population (see Table~S1). A possible explanation is that \ch{BzTr+} is the primary dissociating species, while the other isomers, which absorb less efficiently at this wavelength, may undergo isomerization to \ch{BzTr+} and not direct dissociation. Under this hypothesis (referred to as Hypothesis 2 in the SI), we fitted the kinetic curve at 266~nm using Eq.~S3, assuming an initial \ch{BzTr+} population of 20\%. 

In summary, the analysis of the photoprocessing kinetic curves demonstrates that photoprocessing significantly alters the isomer populations. In particular, it provides clear evidence for interconversion between 1-\ch{NpCH2+} and 2-\ch{NpCH2+}, as well as interconversion --\textit{a priori}-- between the two \ch{NpCH2+} isomers and \ch{BzTr+}.

 \begin{table}[h]
    \centering
     \begin{tabular}{!{\vrule width 2pt}c!{\vrule width 2pt}c|c|c|c|c|c!{\vrule width 2pt}}
     \noalign{\hrule height 2pt}
       $\lambda~\rm(nm)$   & \( k_{1} \) & \( A_{1} \) & \( k_{2} \) & \( A_{2} \) & \( k_{3} \) & \( A_{3} \) \\
      \noalign{\hrule height 2pt}
       \multirow{2}{*}{564}  & 0.141 &  0.12 & 0.03 & 0.43 & 0 & 0.43 \\
       \cline{2-7}
        & \multicolumn{4}{c|}{{\it{1-\ch{NpCH2+}}} + {\it{2-\ch{NpCH2+}}}}   & \multicolumn{2}{c!{\vrule width 2pt}}{{\it{\ch{BzTr+}}}} \\
       \hline 
       \multirow{2}{*}{548}  & 0.27  &  0.35 & 0.03 & 0.28 & 0 & 0.38 \\
       \cline{2-7}
        & \multicolumn{4}{c|}{{\it{1-\ch{NpCH2+}}} + {\it{2-\ch{NpCH2+}}}}   & \multicolumn{2}{c!{\vrule width 2pt}}{{\it{\ch{BzTr+}}}} \\
       \hline 
       \multirow{2}{*}{538}  & 0.172 &  0.29 & 0.0124 & 0.33 & 0 & 0.38\\
       \cline{2-7}
        & \multicolumn{4}{c|}{{\it{1-\ch{NpCH2+}}} + {\it{2-\ch{NpCH2+}}}}   & \multicolumn{2}{c!{\vrule width 2pt}}{{\it{\ch{BzTr+}}}} \\
       \hline 
       \multirow{2}{*}{516}  & 0.346 &  0.38 & 0.011 & 0.35 & 0 & 0.27\\
       \cline{2-7}
        & \multicolumn{2}{c|}{{\it{1-\ch{NpCH2+}}}}  & \multicolumn{2}{c|}{{\it{2-\ch{NpCH2+}}}}   & \multicolumn{2}{c!{\vrule width 2pt}}{{\it{\ch{BzTr+}}}} \\
       \hline 
       \multirow{2}{*}{480}  & 0.349 &  0.36 & 0.00679 & 0.21 & 0.000482 & 0.42\\
       \cline{2-7}
        & \multicolumn{2}{c|}{{\it{1-\ch{NpCH2+}}}}  & \multicolumn{2}{c|}{{\it{\ch{BzTr+}}}}   & \multicolumn{2}{c!{\vrule width 2pt}}{{\it{2-\ch{NpCH2+}}}} \\
       \hline 
       \multirow{2}{*}{454}  & 0.265 &  0.31 & 0.015 & 0.14 & 0.00106 & 0.55\\
       \cline{2-7}
        & \multicolumn{4}{c|}{{\it{1-\ch{NpCH2+}}} + {\it{\ch{BzTr+}}}}   & \multicolumn{2}{c!{\vrule width 2pt}}{{\it{2-\ch{NpCH2+}}}} \\
       \hline 
       \multirow{2}{*}{444}  & 0.272 &  0.2 & 0.0174 & 0.22 & 0.00128 & 0.58\\
       \cline{2-7}
        & \multicolumn{4}{c|}{{\it{1-\ch{NpCH2+}}} + {\it{\ch{BzTr+}}}}   & \multicolumn{2}{c!{\vrule width 2pt}}{{\it{2-\ch{NpCH2+}}}} \\  
      \noalign{\hrule height 2pt}
     \end{tabular}
         \caption{Parameters derived from the three-exponential decay model used to fit the photo-processing kinetic curves presented in Figure~\ref{PCK1MeNp}. The $k$ rates are given in $\rm s^{-1}$. Tentative assignment of the different parameters to specific isomers.}
     \label{tablePKC_1-MeNp}
 \end{table}

\subsubsection{Direct evidence for interconversion between isomers}
\label{Interconversion}

To better understand how isomer populations evolve with photoprocessing time, we analyzed the temporal behavior of both reactive and non-reactive ion channels. These results are presented in Figure~\ref{reactivityKinetic_log} (logarithmic time scale).

At 444 and 454~nm, the intensity of the non-reactive channel steadily decreases, approaching zero at long timescales. This behavior is consistent with the presence of \ch{BzTr+}, which eventually dissociates. In contrast, the reactive population decays much more slowly, reflecting the weak absorption of 2-\ch{NpCH2+} at these wavelengths, as discussed in the previous subsection. As observed in the 266~nm kinetics, 2-\ch{NpCH2+} likely undergoes slow isomerization into 1-\ch{NpCH2+} or \ch{BzTr+}, which then dissociates more efficiently upon further photon absorption.

At 516~nm, both \ch{NpCH2+} isomers absorb, but \ch{BzTr+} does not. This leads to a marked decrease in the reactive population due to dissociation, while the non-reactive population is expected to remain stable, which is observed up to about 40~s of irradiation. At longer timescales, a rise in the non-reactive channel is observed suggesting that non-reactive species are formed from photoprocessing of the reactive species. This further supports the previous suggestion that \ch{NpCH2+} isomers convert into \ch{BzTr+}.

The behavior observed at 480~nm appears intermediate: the reactive population decays similarly to that at 444 and 454~nm, while the non-reactive channel shows trends more akin to those at 516 and 564~nm. This illustrates the influence of the irradiation wavelength on the temporal evolution of the isomer populations and the complexity to analyse these results.

\begin{figure}[!ht]
    \centering
    \includegraphics[width=1\linewidth]{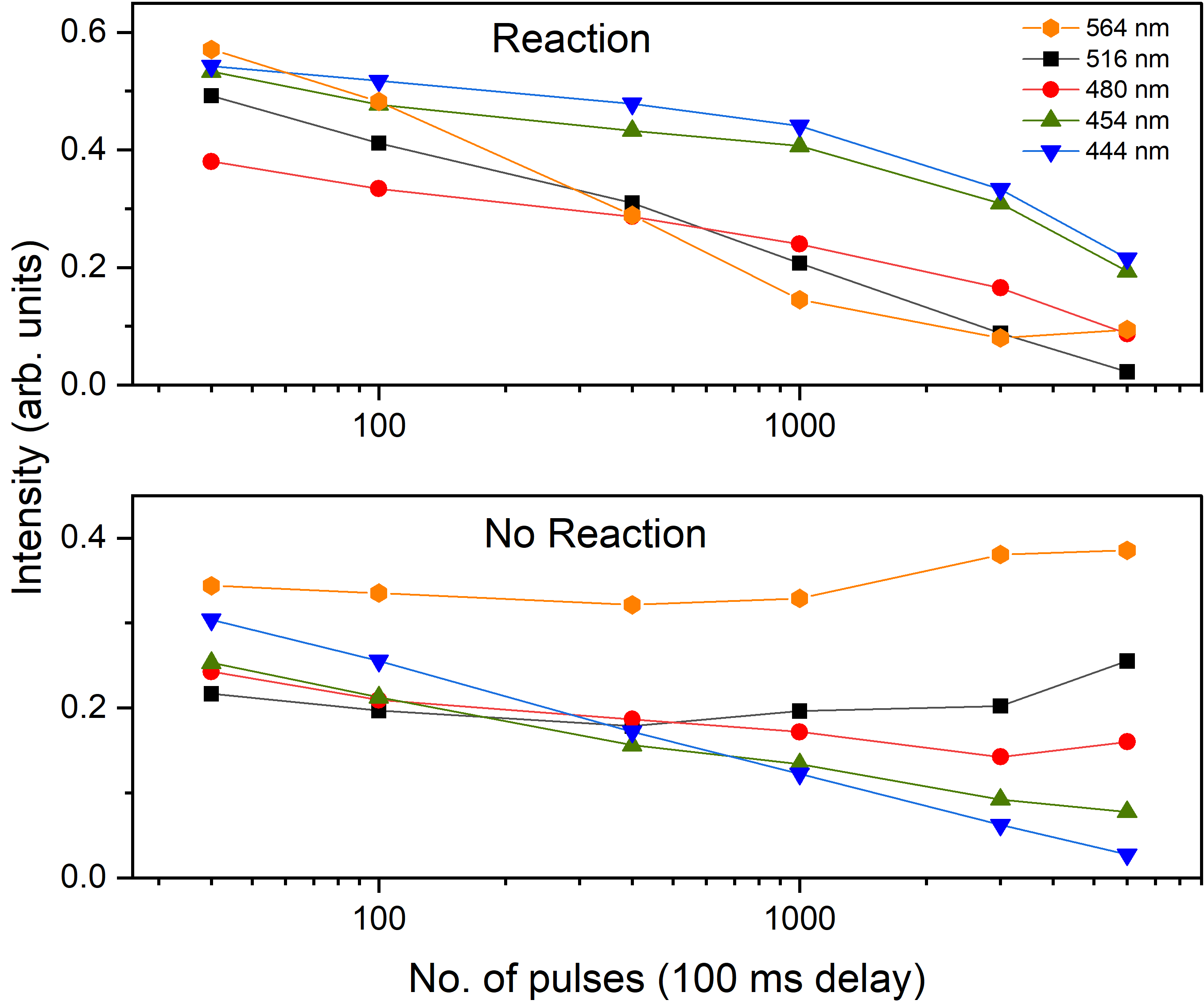}
    \caption{Relative populations of the reactive (1- and 2-\ch{NpCH2+}) and non-reactive (\ch{BzTr+}) species as a function of irradiation time at 444, 454, 480, 516, and 564~nm. The remaining populations after irradiation were measured through reactivity between \ch{C11H9+} ions and their neutral precursor, 1-MeNp (see text for further details). Note that the intensities are not normalized but they are corrected at best for ion current variations by using the decays given by the photoprocessing kinetic curves presented in Figure~\ref{PCK1MeNp}.}
    \label{reactivityKinetic_log}
\end{figure}

To conclude, we report evidence for the interconversion of the \ch{NpCH2+} isomers into \ch{BzTr+}, which can occur for energies below the dissociation threshold. However, we find no evidence in our experiments for a regime in which the reverse reaction is observed.

%%%%%%%%Theoretical results%%%%%%%%%%
\subsection{Results of MD/DFTB simulations}
\label{MD}

The branching ratios (BRs) for the dissociation of the two cationic precursors, 1- and 2-\ch{MeNp^{.}+}, leading to the formation of \ch{C11H9+} species, are reported in Table~\ref{tab:BR_MD}. It is important to note that these BRs indicate general trends rather than absolute values, as the dissociation efficiency is too low to yield statistically significant data. As expected, fragmentation efficiency increases with energy and remains similar for both 1- and 2-\ch{MeNp^{.}+}. At the two examined energies (13.0 and 14.0 eV), hydrogen loss (\ch{H}) is the dominant dissociation pathway, with comparable BRs relative to the total number of fragments for both precursors. However, the simulations also reveal substantial hydrocarbon loss, primarily through the formation of \ch{C2H2} and \ch{C2H3}. This latter pathway is not observed experimentally and is predicted to decrease in the simulations with lower internal energy, dropping from 46\% at 14 eV to 29\% at 13 eV for both isomers.

\begin{table}[!ht]
    \centering
    \begin{tabular}{|c||c|c||c|c|}
    \hline
 Internal energies (eV) &\multicolumn{2}{c ||}{13.0}  & \multicolumn{2}{c|}{14.0}  \\ \hline
     Precursor   & 1-\ch{MeNp^{.}+} & 2-\ch{MeNp^{.}+} & 1-\ch{MeNp^{.}+} & 2-\ch{MeNp^{.}+}\\ \hline
          \#frags./\#simuls. & 6 &  3 & 18 & 16 \\ 
        \#\ch{C11H9+}/\#frags. & 73 & 71 & 54 & 54 \\
        \#1-\ch{NpCH2+}/\#\ch{C11H9+}&  63 & 20 & 59 & 20 \\
        \#2-\ch{NpCH2+}/\#\ch{C11H9+} & 3 & 40 & 17 & 43 \\
        \#\ch{BzTr+}/\#\ch{C11H9+} & 33 & 40 & 21 & 26 \\
        \#\ch{C5C6+}/\#\ch{C11H9+} & - & - & 3 & 11\\ \hline
    \end{tabular}
        \caption{Branching ratios (in \%) for the dissociation of 1- and 2-\ch{MeNp^{.}+}, obtained from MD/DFTB simulations over a 1 ns timescale at two different internal energies. The values are averaged over 720 simulations with randomly assigned velocity distributions. Note: \# denotes the number of occurrences.}
    \label{tab:BR_MD}
\end{table}

Regarding the nature of the isomers resulting from H loss, all three isomers identified experimentally, i.e., 1-\ch{NpCH2+}, 2-\ch{NpCH2+} and \ch{BzTr+} are observed at the two energies for both precursors. While only these three isomers form at 13 eV, additional species containing both 5- and 6-membered rings appear at 14 eV.

The following trends can be observed from the abundance of each isomer as a function of internal energy and precursor. For the two considered energies, 1-\ch{NpCH2+} is the most abundant isomer when fragmenting 1-\ch{MeNp^{.}+}, whereas 2-\ch{NpCH2+} dominates for 2-\ch{MeNp^{.}+}. 
At the lowest energy, the BR for  2-\ch{NpCH2+} from 1-\ch{MeNp^{.}+} is very small whereas the BR for 1-\ch{NpCH2+} is half that of 2-\ch{NpCH2+} from 2-\ch{MeNp^{.}+}.

Besides, the abundances of both \ch{NpCH2+} isomers show no significant dependence on the deposited energy when fragmenting 2-\ch{MeNp^{.}+}. In contrast, for 1-\ch{MeNp^{.}+}, the number of 2-\ch{NpCH2+} fragments relative to that of 1-\ch{NpCH2+} increases with energy. Finally, for both precursors, the BRs of \ch{BzTr+} fall between those of the two \ch{NpCH2+} isomers, showing similar values for both precursors and a tendency to decrease with increasing energy. This suggests comparable activation barriers, likely due to the rate-limiting step in the formation of \ch{BzTr+} from \ch{MeNp^{.}+}, namely the final H-loss step (see Figure~7 in ref.\cite{Wenzel2022}). 

We analyzed a few representative trajectories of the MD simulations starting from \ch{MeNp^{.}+} for the lowest probed energy  
(see Figure~\ref{fig:MD_snapshots}). This allows us to get insights into isomerization and dissociation mechanisms leading to the formation of the three isomers. (1)  The formation of 1-\ch{NpCH2+} (resp. 2-\ch{NpCH2+}) from 1-\ch{MeNp^{.}+} (resp. 2-\ch{MeNp^{.}+})  proceeds through H migration from the -\ch{CH3} group to one carbon of the aromatic cycle that becomes \( sp^{3} \) (step 1$\rightarrow$2 in Figure~\ref{fig:MD_snapshots}, panels (a) and (b)).
H loss is observed from this \( sp^{3} \) carbon atom. (2) The formation of \ch{BzTr+} typically occurs via a mechanism involving intermediates similar to 1 to 4 drawn in panel (b); H migration from the \ch{CH3} group to the carbon ring is followed by insertion of the -\ch{CH2} group into the 6-membered ring. The H loss occurs from an intermediate such as \#4 or \#5 of panel (b) in Figure~\ref{fig:MD_snapshots} to lead to \ch{BzTr+}. (3) Several pathways were identified for the formation of 2-\ch{NpCH2+} (resp. 1-\ch{NpCH2+}) from 1-\ch{MeNp^{.}+} (resp. 2-\ch{MeNp^{.}+}). It  proceeds through H migration  either followed by \ch{CH2} migration through the formation of a 3-carbon ring intermediate (panel (a))  or followed by \ch{CH2} insertion into a 7-carbon ring (steps \#1 to \#5 of panel (b)) and then \ch{CH2} migration to another carbon atom  (step \#5 to \#6 of panel (b)). Interestingly, in all cases, isomerization occurs prior to dissociation and the final H loss arises from a \( sp^{3} \) carbon atom.

\begin{figure}[!htbp]
\centering
   \includegraphics[width=9cm]{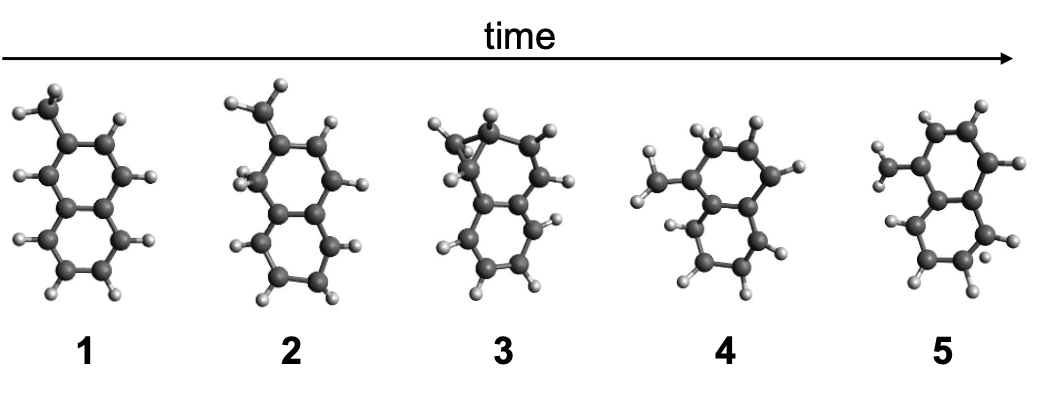} \\
    (a) \\
    \vspace{0.3cm}

   \includegraphics[width=9cm]{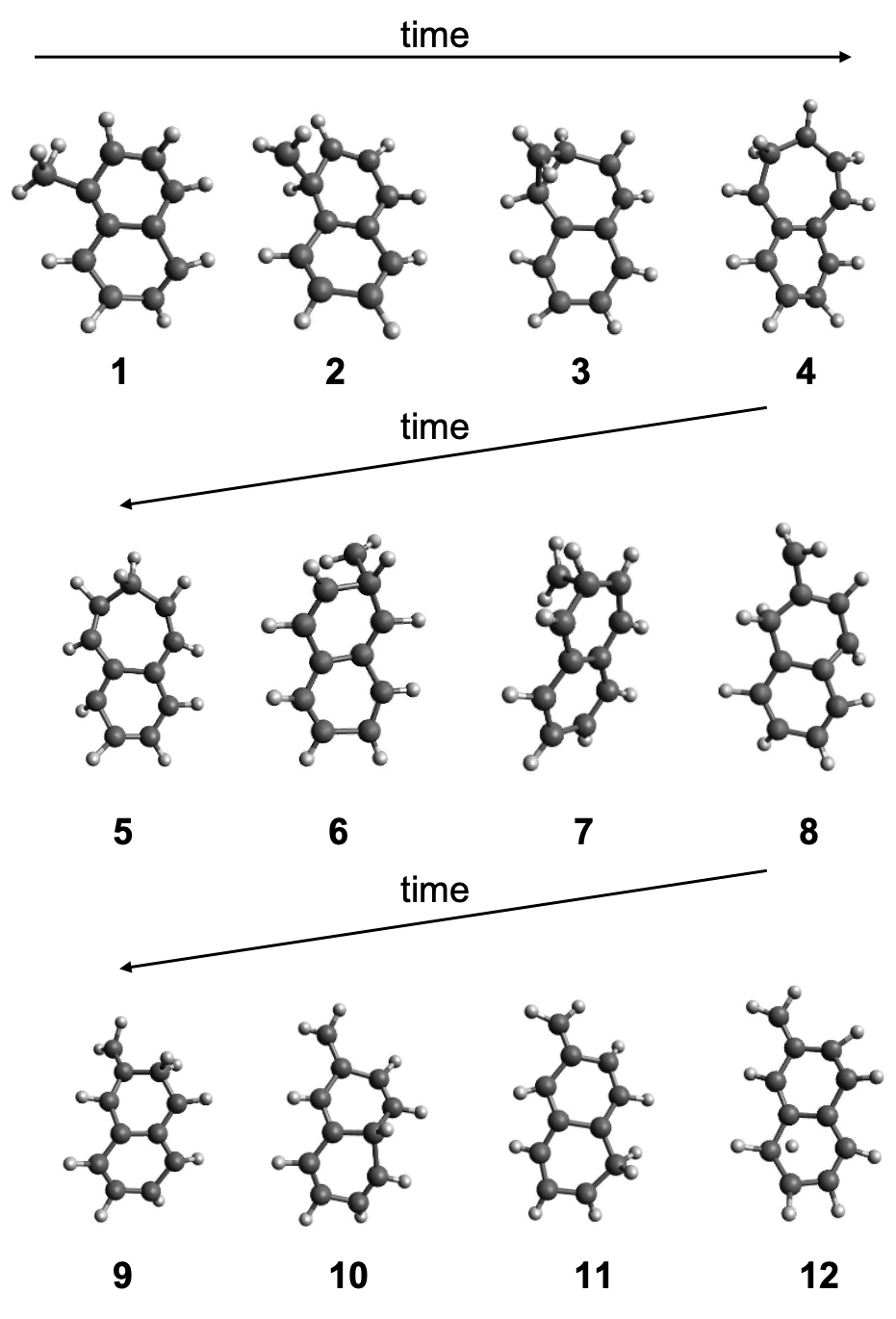} \\
    (b) \\
  \caption{Snapshots taken from two independent simulations at 13 eV for the dissociation of (a) 2-\ch{MeNp^{.}+} resulting in 1-\ch{NpCH2+} and (b) 1-\ch{MeNp^{.}+} resulting in 2-\ch{NpCH2+}. These dissociation mechanisms are shown for a given precursor; however, based on our studies, neither mechanism can be excluded for the other precursor.}
  \label{fig:MD_snapshots}
\end{figure}

\begin{figure}[!htbp]
\centering
\includegraphics[width=8.5cm]{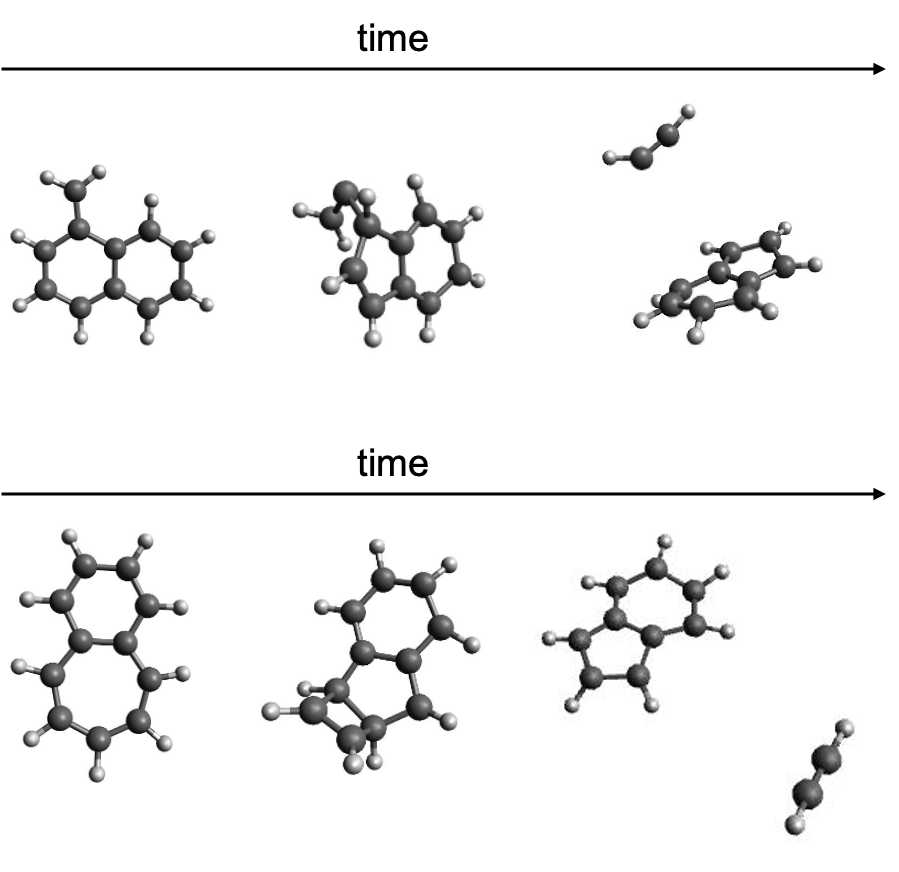} \\
    
  \caption{Snapshots taken from two independent simulations at 11.6 eV for the dissociation of 1-\ch{NpCH2+} (top) and \ch{BzTr+} (bottom).}
  \label{fig:snapshots}
\end{figure}

Simulations were also performed to investigate the dissociation paths of the three isomers of \ch{C11H9+} identified experimentally (1-\ch{NpCH2+}, 2-\ch{NpCH2+}, and \ch{BzTr+}). The lowest energy at which dissociation was observed (few dissociation events during the 1~ns timescale of the simulations) is 11.6~eV. All three isomers dissociate at this energy, exclusively undergoing \ch{C2H2} loss to form the indenyl cation. More specifically, among 180 simulations conducted for each isomer, 1, 3, and 5 dissociation events were observed for 1-\ch{NpCH2+}, 2-\ch{NpCH2+}, and \ch{BzTr+}, respectively.
 Snapshots from 1-\ch{NpCH2+} and \ch{BzTr+} dissociation trajectories at 11.6~eV are reported in Figure~\ref{fig:snapshots}. The loss of \ch{C2H2} from 1-\ch{NpCH2+} was found to occur via a fairly straightforward mechanism with only one visible intermediate containing a 5-membered ring and a \ch{CCH2} group. H migration and loss of \ch{C2H2} follow. The time between the printed intermediates (250 fs) is too short to visualize all H migrations. The loss of \ch{C2H2} from \ch{BzTr+} is typically preceded by the formation of a 5- and 4-membered ring prior to dissociation of \ch{C2H2}.  In the case of 2-\ch{NpCH2+}, the mechanisms leading to the formation of the 5-membered ring were found to be different in the three simulations. In one case, the mechanism is similar to that observed in 1-\ch{NpCH2+}. In the second one, H migration and ring opening are observed. In the third one, there is first isomerization into the \ch{BzTr+} and then the  dissociation mechanism is similar to that of \ch{BzTr+} reported in Figure~\ref{fig:snapshots}. In conclusion, the loss of neutral acetylene from the three isomers was found to systematically proceed through the formation of a 5-carbon ring intermediate with a grafted [2C,2H] group that can be \ch{CCH2} (top panel in Figure~\ref{fig:snapshots}) or \ch{C2H2} (bottom panel in Figure~\ref{fig:snapshots}).
 
From our simulations, isomerization into \ch{BzTr+} prior to dissociation is also possible but not systematic.
As mentioned in the previous paragraph, the fragmentation efficiency remains low at 11.6 eV. Interestingly, among the simulations that do not result in fragmentation, a small fraction leads to isomerization. Specifically, 7\% of \ch{BzTr+} isomerizes into either 1-\ch{NpCH2+} or 2-\ch{NpCH2+}. Conversely, 4\% of 2-\ch{NpCH2+} and 3\% of 1-\ch{NpCH2+} isomerize into \ch{BzTr+}. Finally, 3\% of 1-\ch{NpCH2+} isomerizes into 2-\ch{NpCH2+} while 8\% of 2-\ch{NpCH2+} isomerizes into 1-\ch{NpCH2+}. This isomerization efficiency increases when the internal energy increases for the two isomers. These isomerization pathways are not the only ones : the formation of isomers possessing 5-carbon rings is also observed.

When increasing the energy to 12.5\,eV, the fragmentation efficiency increases (9, 14 and 7 losses of \ch{C2H2} in the cases of 1-\ch{NpCH2+}, 2-\ch{NpCH2+} and \ch{BzTr+}, respectively). The indenyl cation remains the predominant fragment; however, during the dissociation of 1- and 2-\ch{NpCH2+}, additional \ch{C9H7+} isomers featuring a single six-membered ring with attached carbon chains are also formed.
Finally, other fragmentation paths such as the loss of \ch{C4H2} start to occur.

%%%%%%%%Discussion%%%%%%%%%%
\section{Discussion}
\label{Discussion}

In the following, we present a comprehensive discussion on the long-lived isomers of the -H fragment of MeNp cation, as well as further considerations into their isomerization. Moreover, a final section is dedicated to the astrophysical implications of our findings.

%%%%%%%%Long-lived isomers%%%%%%%%%%
\subsection{Long-lived isomers of \ch{C11H9+} produced from MeNp}
\label{Long-lived isomers}

Nagy {\it et al.} \cite{Nagy2011} identified 2-\ch{NpCH2+} and \ch{BzTr+} by recording the electronic photo-absorption spectrum of \ch{C11H9+} ions generated from 1-MeNp in a hot-cathode discharge source and isolated in Ne matrices. Wenzel {\it et al.} \cite{Wenzel2022} investigated the long-lived isomers of gas-phase \ch{C11H9+} using infrared predissociation (IRPD) spectroscopy of trapped Ne-tagged \ch{C11H9+} species, combined with saturation depletion measurements at selected vibrational bands. In their experiments, \ch{C11H9+} ions were produced through collisions between 17~eV electrons and 2-MeNp. The authors determined a maximum relative abundance of 40\% for \ch{BzTr+}.
Regarding the two \ch{NpCH2+} isomers, their investigation was limited to the 2-\ch{NpCH2+} isomer, and as such, they were unable to confirm or exclude the presence of the 1-\ch{NpCH2+} isomer.

From our experiments, we were able to confirm the presence of the two isomers identified by Nagy {\it et al.} \cite{Nagy2011} and Wenzel {\it et al.} \cite{Wenzel2022}, namely \ch{BzTr+} and 2-\ch{NpCH2+}. Additionally, we revealed a broad feature in the [460-530] nm range of the MPD spectra that was not observed in the photo-absorption spectrum \cite{Nagy2011}. We attributed this feature to 1-\ch{NpCH2+}, which suggests the presence of two relatively intense bands within this broad feature, based on the calculations reported by Nagy {\it et al.} \cite{Nagy2011} and our own calculations (see Table~S3). 

In Section~\ref{PKCreactivity}, we determined a relative abundance of 20\% for \ch{BzTr+}, but we found that it can be up to ~50\% when using the Xe lamp and minimizing photo-processing by 266~nm photons.
We derived the relative abundance of 1-\ch{NpCH2+} by analyzing photoprocessing kinetic curves and concluded that this abundance is comparable to that of 2-\ch{NpCH2+}, typically 40\%. Because of interconversion processes, it is however very likely that the relative initial abundance of 1-\ch{NpCH2+} is significantly higher than that of 2-\ch{NpCH2+} in our experiment. This is in line with MD simulations predicting a significantly higher abundance of 1-\ch{NpCH2+} relative to 2-\ch{NpCH2+} when 1-MeNp is used as the precursor.

The non-identification of 1-\ch{NpCH2+} in previous studies may therefore stem from photoprocessing effects that lead to its preferential isomerization or dissociation. Alternatively, it could be due to differences in the production method of \ch{C11H9+}, particularly the use of electrons rather than photons (at 266~nm).
In the experiment by  Nagy {\it et al.} \cite{Nagy2011}, the authors inferred the presence of 1-\ch{NpCH2^{.}} radicals after photoprocessing the Ne matrix with a medium-pressure mercury lamp (+~water filter).
These observations suggest that 1-\ch{NpCH2+} was likely present in their experiments, but was somehow lost during their measurements. Their spectroscopic method involved exposing the Ne matrix to broadband irradiation from a high-pressure xenon light source before spectral analysis, leading to continuous photoprocessing of the trapped \ch{C11H9+} ions. Our experiments demonstrate that this should result in preferential photodissociation of 1-\ch{NpCH2+}. This provides a likely explanation for why \citet{Nagy2011} were unable to observe its absorption spectrum, despite it being expected to have the strongest photo-absorption bands among the three low-energy \ch{C11H9+} isomers in the [430-590]~nm spectral range.

Wenzel {\it et al.} \cite{Wenzel2022} considered only \ch{BzTr+} and 2-\ch{NpCH2+} in their analysis of the recorded IRPD spectrum. However, they noted that their synthetic spectrum did not fully account for all observed experimental bands, suggesting the possible presence of additional isomers. To further investigate the contributions of 1-\ch{NpCH2+} in their experiments, we calculated its anharmonic IR spectra at the same theoretical level used by Wenzel {\it et al.} \cite{Wenzel2022}. Geometries were first optimized at the density functional theory (DFT) level using the hybrid functional B3LYP  in conjunction with the def2TZVPP~\citep{weigend2005} basis set. Anharmonic spectra were computed using the perturbative approach (VPT2)~\cite{barone2005}. These calculations were performed with the Gaussian16 suite of programs~\cite{g16}.

Figure~\ref{IRspectra} compares the experimental IRPD spectrum from Wenzel {\it et al.} \cite{Wenzel2022} with both their original synthetic spectrum (middle panel of Figure~\ref{IRspectra}) and our newly generated synthetic spectrum (top panel of Figure~\ref{IRspectra}), which assumes a mixture of 45\% 1-\ch{NpCH2+}, 25\% 2-\ch{NpCH2+}, and 30\% \ch{BzTr+}. This composition significantly improves the agreement between the synthetic and experimental spectra. Specifically, it accounts for an additional band at 770~\ch{cm^{-1}} in the out-of-plane CH bending ($\gamma_{CH}$) region, as well as a new feature at 1336~\ch{cm^{-1}} in the in-plane CH bending ($\delta_{CH}$) and CC stretching ($\nu_{CC}$) region. Furthermore, its inclusion reduces the intensity of the $\nu_{CC}$ band at 1636~\ch{cm^{-1}}, which was previously overestimated in the synthetic spectrum. The obtained spectral agreement allowed us to firmly identify 1-\ch{NpCH2+} in this previous investigation. 

\begin{figure}[!ht]
    \centering
    \includegraphics[width=1\linewidth]{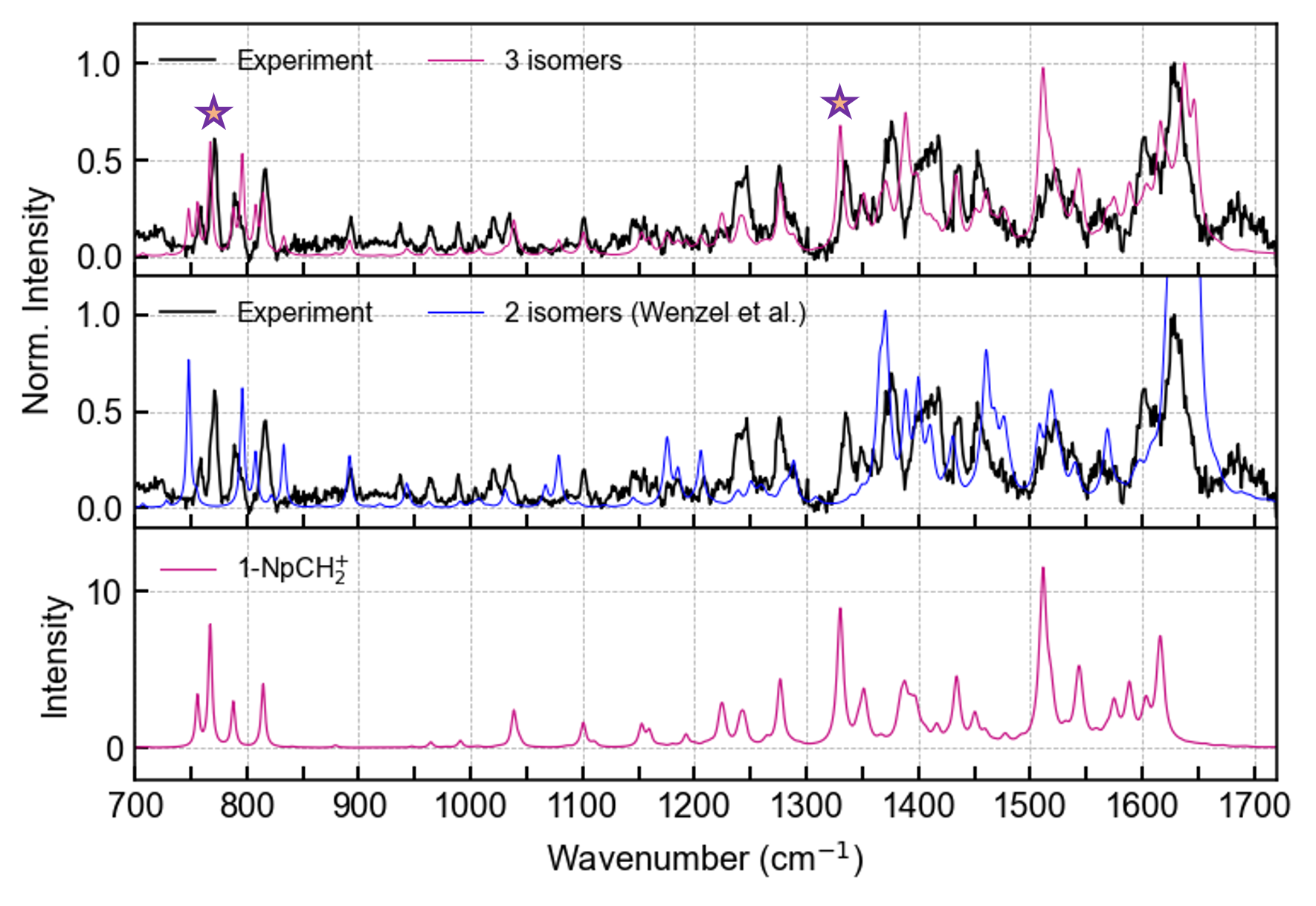}
    \caption{Experimental IRPD spectrum of Ne-tagged \ch{C11H9+} (black), as reported by Wenzel {\it et al.}\cite{Wenzel2022}, compared to synthetic spectra generated from the anharmonic vibrational spectra of different isomers. Top panel (pink): 1-\ch{NpCH2+} (0.45), 2-\ch{NpCH2+} (0.25), and \ch{BzTr+} (0.30). The pink stars indicate key new bands at 770 and 1336 $\rm{cm^{-1}}$.  Middle panel (blue): 2-\ch{NpCH2+} (0.70), and \ch{BzTr+} (0.30). Bottom panel: Calculated anharmonic IR spectra of 1-\ch{NpCH2+}. Note that the numerical values in parenthesis indicate the relative abundances used to generate each synthetic spectrum. See text for further details.}
    \label{IRspectra}
\end{figure}

Figure~\ref{IRspectra} shows the possibility to accommodate more isomer diversity. We therefore investigated the possible presence of an additional isomer that was also formed in the MD simulations, i.e. 1,2-\ch{inde(CH2)2+} (see Figure~S6 in the SI). The modifications of the spectrum are not as drastic as those induced by the presence of 1-\ch{NpCH2+} but the fraction of 1,2-\ch{inde(CH2)2+} is also the smallest one (15\%). Still, the relative intensity ratio of bands in the 1300-1700~\ch{cm^{-1}} region is improved as well as the shape of the 1600~\ch{cm^{-1}} band. This can be explained by the very intense band at 1637~\ch{cm^{-1}} in the unscaled harmonic spectrum of 1,2-\ch{inde(CH2)2+}, lower in energy than the most intense bands at 1655 and 1663~\ch{cm^{-1}} in the harmonic spectra of 1-\ch{NpCH2+} and 2-\ch{NpCH2+}, respectively.

In conclusion, all these results demonstrate that various formation mechanisms, including dissociative ionization by electron bombardment and sequential photoionization followed by photodissociation, can generate the three long-lived isomers -- 1-\ch{NpCH2+}, 2-\ch{NpCH2+}, and \ch{BzTr+} -- in comparable abundances, regardless of whether the precursor is 1-MeNp or 2-MeNp.

%%%%%%%%Dissociation and Isomerization???%%%%%%%%%%
\subsection{Competition between isomerization and dissociation}
\label{DissIsom}

Our results provide clear evidence of interconversion between isomers. These changes are most apparent on long timescales (see Figure~\ref{reactivityKinetic_log}), giving support to the fact that isomerization arises from populations carrying a significant internal energy due to gradual build-up of this internal energy upon OPO irradiation.

As a first approximation, we can assume that the dissociation of \ch{C11H9+}requires a similar energy as that of the naphthalene cation \ch{C10H8+}, which has been determined to be around 6.7~eV for a dissociation rate of 10$^2$~s$^{-1}$ of astrophysical relevance \cite{Jochims1999}.
Our MPD spectra cover the 430–590 nm range, corresponding to photon energies between 2.10 and 2.88 eV. Given that at 300~K, ions carry approximately 0.2 eV of internal energy, we conclude that three photons are required for \ch{C11H9+} to dissociate. The probability of absorbing three photons within the same pulse strongly depends on the laser pulse energy and the absorption cross-section \cite{Useli-Bacchitta2010}. 
There is therefore a threshold in absorption cross-section below which dissociation in a single pulse will not be achieved. In this case, the dissociation becomes possible through accumulation of  internal energy  across multiple pulses. The resulting heating rate is modulated by a radiative cooling rate, which leads to an internal energy distribution \cite{Boissel1997}. In this regime, dissociation can occur upon absorbing a single photon, provided the ion has already gained sufficient internal energy \cite{Boissel1997, Martin2015}. 
The recorded MPD spectrum corresponds then to that of the hot ion population, which is expected to be significantly broadened by temperature effect. In addition, if isomerization occurs, it means that the isomer that will ultimately dissociate is different from the initial isomer. This breaks a classical MPD scheme in which one assumes that the observed spectrum somehow reflects the absorption of the first photon. In this context, the discussion on the parameters $A$ and $k$ for photoprocessing kinetics curves (Table~\ref{tablePKC_1-MeNp} and Section~\ref{Photoprocessing}) becomes an oversimplification. Indeed, these parameters likely reflect both the dissociation and isomerization dynamics, in a regime where each isomer has a distinct internal energy distribution due to differential coupling with the radiation field. Consequently, the experimental results cannot be directly compared to MD simulations, where all isomers are assumed to possess the same internal energy (see Section~\ref{MD}).
 
Our results provide clear evidence of isomer interconversion. First, in the analysis of the photoprocessing kinetic curves, we reported evidence for  interconversion of 1-\ch{NpCH2+} into  2-\ch{NpCH2+} (section~\ref{Photoprocessing}), and of 1-\ch{NpCH2+} and/or 2-\ch{NpCH2+} into \ch{BzTr+} (section~\ref{Interconversion}).
Second, additional evidence for isomerization can be found in the MPD spectra. The spectrum of 2-\ch{NpCH2+} (Figure~\ref{spectraisomers}(b)) exhibits an underlying continuum which can be attributed to hot 1-\ch{NpCH2+}. This suggests that 2-\ch{NpCH2+} can isomerize into  1-\ch{NpCH2+}. In contrast, the spectrum of \ch{BzTr+} (Figure~\ref{spectraisomers}(a)) shows no spectral signatures attributable to 1-\ch{NpCH2+}, suggesting that interconversion from \ch{BzTr+} to 1-\ch{NpCH2+} is unlikely under the given conditions. Furthermore, Figure~\ref{MPDspectrum} illustrates how the shape of the broad absorption feature evolves with varying OPO irradiation conditions. The observed flattening of the spectrum for longer irradiation times, along with increased signal in the spectral regions corresponding to \ch{BzTr+} and 2-\ch{NpCH2+}, is most plausibly explained by the isomerization of hot 1-\ch{NpCH2+} into these other isomers, which then dissociate upon further photon absorption.

Determining an isomerization barrier is out of reach for this study.
Previous determination of a barrier for similar systems are all based on theory. A value of around 3~eV was determined for  conversion of the benzylium cation into the tropylium cation \cite{Smith1997, Bullins2011, Fridgen2004, Ignatyev2000} and of around 3.7~eV from the conversion of naphthalene cation into azulene cation \cite{Dyakov2006,Lee2023}. The lowest barrier for \ch{C2H2} loss from the benzylium isomer has been theoretically estimated to be around 3.97~eV \cite{Zhao2017}. Slightly higher values have been calculated for naphthalene (4.75~eV) and azulene (4.72~eV) cations \cite{Dyakov2006,Lee2023}. Rapacioli {\it et al.} .\cite{Rapacioli2015} have determined four pathways with similar barrier heights (3.5-4~eV) between the two isomers, benzylium-like (1-\ch{PyCH2+}) and tropylium-like (\ch{PyTr+}), of the -H fragment of methyl-pyrene cation (\ch{C17H11+}). These results support our finding that the conversion of 1-\ch{NpCH2+} and/or 2-\ch{NpCH2+} into \ch{BzTr+} can occur below the dissociation threshold. Our work aligns with the study by Fridgen {\it et al.}\cite{Fridgen2004}, in which the authors proposed that isomerization into \ch{Tr+} competes with the radiative cooling of excited \ch{Bz+}. The irradiation conditions in our experiments indeed generate vibrationally hot \ch{NpCH2+} populations, favoring isomerization to \ch{BzTr+}. In contrast, the reverse process is not observed, suggesting either a higher barrier for isomerization or more efficient cooling of \ch{BzTr+}.

%%%%%%%%Astro implications%%%%%%%%%%%%

\subsection{Astrophysical implications}
\label{AstroRelevance}

In photodissociation regions exposed to strong UV irradiation, gas-phase \ch{C11H9+} ions are expected to undergo rapid photodissociation, leading to short lifetimes -- unless they reside in UV-shielded or mildly irradiated environments. Therefore, searching for these ions in dark clouds, such as TMC-1, could be more promising, provided they are formed in sufficient quantities. Under mild UV/visible irradiation, our experiments show that the tropylium form will be favored. For harder UV photons, \ch{C11H9+} ions dissociate to form the indenyl cation (\ch{C9H7+}), which could also be a relevant ion to search for in astrophysical environments such as TMC-1. Similar to other PAHs, methylnaphthalene is expected to be primarily condensed on grains (see discussion in Wenzel {\it et al.} \cite{Wenzel2024b}) and released into the gas phase upon energetic processing by cosmic rays or UV photons. This desorption process may or may not be accompanied by dissociative ionization. In the gas phase, \ch{C11H9+} can be formed from methylnaphthalene upon absorbing a single UV photon with an energy above 11.8~eV \cite{Jochims1999}. 

Once formed, \ch{C11H9+} can undergo dissociative recombination with electrons, leading to products with internal energies of approximately 6.7~eV for \ch{NpCH2^{.}} and 6~eV for \ch{BzTr^{.}} (see Table~S2). These values are near or below the estimated dissociation threshold of their respective ions, suggesting that \ch{BzTr^{.}} and possibly \ch{NpCH2^{.}} radicals could be produced. Notably, Pino {\it et al.} \cite{Pino02} reported the formation of the tropyl radical via electron recombination of tropylium in plasmas. Given their resonance stabilization, these radicals and their cations are relatively unreactive, potentially allowing for significant accumulation in various astrophysical environments, including TMC-1.

The three isomers 1-\ch{NpCH2+}, 2-\ch{NpCH2+}, and \ch{BzTr+} are closed-shell species. Their rotational spectra are therefore expected to be simpler than those of their neutral counterparts, facilitating identification. Moreover, our calculations indicate that \ch{BzTr+} has a dipole moment of 1.2~D, while 1-\ch{NpCH2+} and 2-\ch{NpCH2+} exhibit higher values of 2.0~D and 2.5~D, respectively. These values are typically about twice as large as that of indene (c-\ch{C9H8}) \cite{Cernicharo2021}, enhancing their detectability via rotational spectroscopy in astrophysical environments. 

\citet{Rasmussen2023} performed photodissociation action spectroscopy of protonated oxygen-functionalized PAHs (OPAHs) in the UV-visible range. They reported that these species can exhibit very broad absorption bands extending up to 700~nm, leading them to propose OPAHs as promising candidates for the extended red emission (ERE), whose carriers remain unidentified. We note that \ch{C11H9+} ions are also attractive candidates to contribute to the ERE. According to \citet{Witt2006}, ERE carriers should fulfill a two-step photophysical scheme. The first step involves carrier creation through absorption of far-UV photons with energies exceeding 10.5~eV, which applies to the formation of \ch{C11H9+} via dissociative ionization of methylnaphthalene. In the second step, excitation by more abundant near-UV/visible photons should lead to emission within the ERE spectral range (see e.g., \citet{Gordon1998, VanWinckel2002}). \citet{Nagy2011} measured fluorescence from 2-\ch{NpCH2+} and \ch{BzTr+}, and we report here indirect evidence for fluorescence from 1-\ch{NpCH2+}. Since 1-\ch{NpCH2+} absorbs very efficiently, it may also fluoresce efficiently and potentially contribute to the ERE, particularly on the blue edge of this spectral feature.

%%%%%%%%%%%%%%%%%%%%%%%%%%%%%%%%%%%%%%%%%%%%%%%%%%%%%%%%%%%%%%%%%%%%%
%% Conclusions
%%%%%%%%%%%%%%%%%%%%%%%%%%%%%%%%%%%%%%%%%%%%%%%%%%%%%%%%%%%%%%%%%%%%%

\section{Conclusions}
\label{conclusions}

We have investigated the long-lived isomers of the -H fragment of 1-MeNp cation (\ch{C11H9+}) produced in gas phase by 266 nm irradiation. Using the FTICR-MS setup PIRENEA, we performed multiple measurements, including MPD spectroscopy, photofragmentation kinetics, and ion-molecule reaction studies. Additionally, the dissociation dynamics of 1- and 2-MeNp cations (\ch{C11H10^{.}+}) and their \ch{C11H9+} fragments  were explored through MD/DFTB simulations.

Previous investigations using 1-MeNp\cite{Nagy2011} and 2-MeNp\cite{Wenzel2022} as precursors to produce \ch{C11H9+} species identified 2-\ch{NpCH2+} and \ch{BzTr+} as specific isomers of the \ch{C11H9+} ion. In our experiments, we confirmed the presence of these two isomers and provided clear spectroscopic evidence for the additional presence of 1-\ch{NpCH2+}. We demonstrated that 1-\ch{NpCH2+} was also present in the experiments by Wenzel {\it et al.} \cite{Wenzel2022} and we offered a plausible explanation for why Nagy {\it et al.} \cite{Nagy2011} were unable to identify this isomer in their photo-absorption spectrum. Therefore, this study provides a more comprehensive picture of the long-lived isomers of \ch{C11H9+}, confirming that the three lowest-energy isomers, namely 1-\ch{NpCH2+}, 2-\ch{NpCH2+}, and \ch{BzTr+} are formed through various mechanisms, regardless of whether the precursor is 1-MeNp or 2-MeNp. We could not find evidence in our results of the presence of additional isomers.

We also provided a quantitative assessment of the relative populations of the \ch{C11H9+} isomers formed. The fraction of \ch{BzTr+} was estimated via ion–molecule reactions with the parent neutral species, yielding a relative abundance that ranged from 20\% to 50\%, depending on the formation conditions. In contrast, determining the relative abundances of the \ch{NpCH2+} isomers is more challenging. Taking advantage of the strong dissociation yield of 1-\ch{NpCH2+} near 500~nm, we estimated its relative abundance to be approximately 40\% -- a value obtained in the case where \ch{BzTr+} accounts for 20\%. However, due to strong experimental evidence for interconversion from 1-\ch{NpCH2+} to 2-\ch{NpCH2+}, this estimate should be considered a lower limit. We note however that our derived isomeric abundances are comparable to those obtained by reanalyzing the IRPD spectrum of Wenzel et al. \cite{Wenzel2022}, despite the different ion formation conditions.

Despite the known limitations of MPD spectroscopy, we were able to report band positions between 430 and 590~nm for the three lowest-energy isomers of \ch{C11H9+} in the gas phase. For the first time, we also identified spectral features attributable to 1-\ch{NpCH2+}. The observed broad and congested structure is consistent with a population of hot 1-\ch{NpCH2+}, resulting from its strong absorption cross section and efficient formation via isomerization of 2-\ch{NpCH2+}. 
The origin of the congested structure peaking at $\sim$490~nm requires further investigation. In addition to temperature effects, it may be associated with a low-frequency vibrational mode linked to the internal rotation of the methylidene group, as previously discussed for \ch{Bz+} \cite{Dryza2012}. 
Further characterization would benefit from spectroscopy at cryogenic temperatures using rare-gas tagging, as applied in previous studies of \ch{Tr+}\cite{Jacovella2020}. Moreover, our study motivates additional investigations of photoisomerization processes of \ch{C11H9+}, using ion mobility combined with IR or visible spectroscopy \cite{Masson2015, Bull2017, Musbat2018, Bansal2023}.

The photo-processing conditions in our experiments produced hot ion populations and a complex interplay between isomerization and dissociation pathways, both of which are highly sensitive to the irradiation wavelength. Our analysis reveals interconversion between 1-\ch{NpCH2+} and 2-\ch{NpCH2+}, as well as conversion to \ch{BzTr+}, with no evidence for the reverse reaction. While MD simulations suggest that all interconversions are energetically accessible, our results indicate that the observed pathways are governed by photophysical conditions. This highlights the potential for tuning the relative isomer populations via selective wavelength excitation. For instance, under irradiation with a Xe lamp at wavelengths longer than 610~nm, the \ch{BzTr+} population could be enriched to 50\% of the total isomer population.

The unique properties of the m/z 141 isomers --particularly \ch{BzTr+}, which tends to form from other isomers under photoprocessing and is not readily converted back-- make them compelling candidates for detection in astrophysical environments such as TMC-1. A comprehensive understanding of their photophysics could also enhance photochemical models describing PAH ion chemistry in Titan’s upper atmosphere \cite{Loison2019}.

Finally, while similar studies could be conducted using 2-MeNp as a precursor, the reported methodology also proves to be a powerful tool for investigating the properties of nitrogen-derived species, as recently demonstrated by Arun \textit{et al.} \cite{Arun2023}. Furthermore, exploring the isomerization of larger and more compact species, such as the -H cation of methylpyrene \cite{Rapacioli2015, Wenzel2022}, is of particular interest, especially in the context of astrophysical environments.

\section*{Acknowledgments}

The work presented here was partially funded by the Agence Nationale pour la Recherche, ANR grant No ANR-21- CE30-0010, SynPAHcool. The running costs for cryogenic fluids, required for the superconducting magnet of the PIRENEA setup, were also supported by recurrent funding for the Nanograin platform at IRAP, as well as by the Thematic Action 'Physique et Chimie du Milieu Interstellaire' (PCMI) of the INSU Programme National 'Astro', with contributions from CNRS Physique, CNRS Chimie, CEA, and CNES.
The authors gratefully acknowledge Loïc Noguès, David Murat and Odile Coeur-Joly for their sustained technical support of the PIRENEA setup over many years.
AS thanks the computing mesocenter CALMIP (“CALcul en MIdi Pyr\'en\'ees”, UAR~3667 of CNRS) for generous allocation of computer resources (project p17002). Finally, the authors sincerely thank the reviewers for their valuable suggestions and for encouraging a deeper exploration of our experimental findings. N. Ben Amor and M. Boggio-Pasqua are also acknowledged for valuable discussions regarding theoretical calculations of the excited electronic states of the \ch{C11H9+} isomers.

\section*{Associated content}

\subsection*{Data Availability Statement}
The dataset associated with this work can be found under 10.5281/zenodo.15316509

% \subsection*{Supporting Information}

% The Supporting Information is available free of charge at

% Figure~S1: mass spectrum after 266~nm irradiation of gaseous 1-MeNp; Figure~S2: Mass spectra illustrating the procedure combining photoprocessing and reactivity; Figure~S3: MPD spectrum of \ch{C11H9+} including the normalized intensities for each observed fragment; Figure~S4: Total ion intensity for MPD spectra; Figure~S5: MPD spectra of \ch{C11H9+} under different OPO conditions; Figure~S6: Experimental IRPD spectrum of Ne-tagged \ch{C11H9+}\cite{Wenzel2022} compared to synthetic spectra generated from the anharmonic vibrational spectra of different isomers; Figure~S7: Photofragmentation kinetic curve of \ch{C11H9+} at 266~nm.

% Table~S1: Relative population of the \ch{BzTr+} isomer under different conditions for {\it m/z} 141 production; Table~S2: Calculated ionization potentials; Table~S3: Computed (TDDFT) excitation energies for the five lowest-energy isomers of \ch{C11H9+} along with their oscillator strengths and comparison with experimental values derived from our study and from Nagy \textit{et al.} \cite{Nagy2011}.

\subsection*{Author Contributions}
A. I. L.: Investigation, Formal analysis, Data curation, Visualization, Writing -- original draft, Writing -- review \& editing. \\
A. B.: Instrumentation, Methodology, Investigation, Validation, Resources. \\
A. S.: Methodology, Investigation, Computation, Resources, Writing -- original draft. \\
C. J.: Conceptualization, Methodology, Investigation, Formal analysis, Resources, Writing -- original draft, Writing -- review \& editing, Supervision, Project administration, Funding acquisition.\\
All authors have read and agreed to the published version of the manuscript.

\providecommand{\latin}[1]{#1}
\makeatletter
\providecommand{\doi}
  {\begingroup\let\do\@makeother\dospecials
  \catcode`\{=1 \catcode`\}=2 \doi@aux}
\providecommand{\doi@aux}[1]{\endgroup\texttt{#1}}
\makeatother
\providecommand*\mcitethebibliography{\thebibliography}
\csname @ifundefined\endcsname{endmcitethebibliography}  {\let\endmcitethebibliography\endthebibliography}{}

\appendix
%\section{Supporting Information}
\includepdf[pages=-]{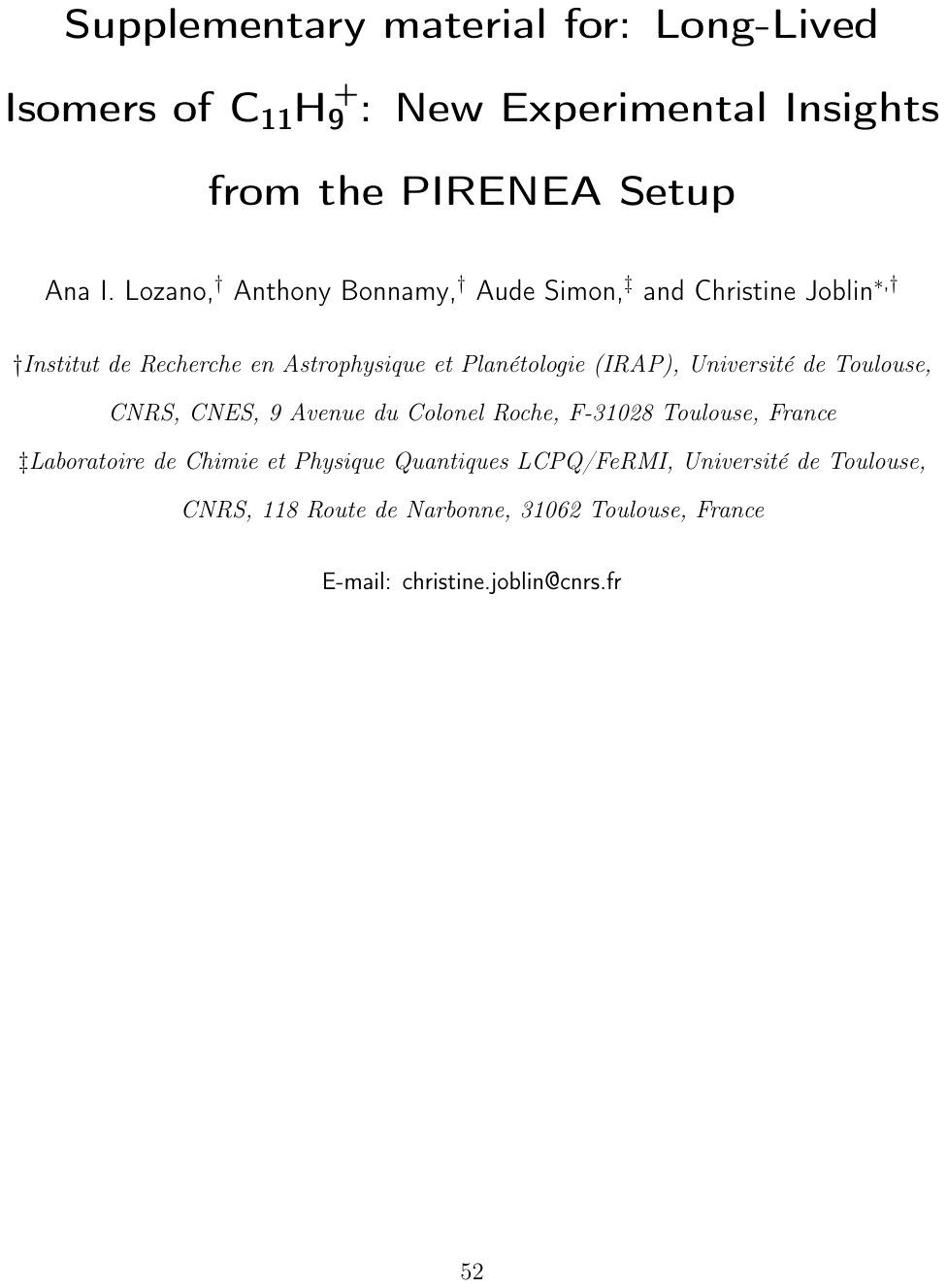}

\end{document}